\documentclass[twocolumn]{autart}    % Enable this line and disable the 
\usepackage{lineno,hyperref}
\usepackage{color}
\usepackage{bm}
\usepackage{epsfig}
\usepackage{pdfpages}
\usepackage{graphicx}
\usepackage{physics}
\usepackage{mathtools}
\usepackage{bbm}
\usepackage{subcaption}
\usepackage{multirow}
\usepackage{amsmath,tabularx} 
\usepackage{enumitem}
\usepackage{mathrsfs}  
\usepackage{pdfpages}
\usepackage{pdflscape}
\usepackage[UKenglish]{babel}
\usepackage{qtree}
\usepackage{rotfloat}
\usepackage{tikz}
\usepackage{tikz-qtree}

\usepackage[xcolor]{changebar}  \cbcolor{blue}
\usepackage{dsfont}
\usepackage{cite}
\usepackage[colorinlistoftodos]{todonotes}
\graphicspath{{Images/}}
\usepackage{bm}
\usepackage{stmaryrd}
\usepackage{mathtools}
\mathtoolsset{showonlyrefs}
\usepackage{amssymb}

\usepackage[acronym]{glossaries}

\newacronym{MA}{MA}{Max-algebraic}
\newacronym{SMPL}{SMPL}{Switching Max-Plus Linear}
\newacronym{MMPS}{MMPS}{Max-Min-Plus-Scaling}

\newcommand{\maxpow}[2]{{#1}^{{\scriptscriptstyle\otimes}^{\scriptstyle{#2}}}} 
\newcommand{\RN}[1]{%
	\textbf{\uppercase\expandafter{\romannumeral#1}}%
}
\newcommand{\rn}[1]{%
	\textit{\lowercase\expandafter{\romannumeral#1})}%
}
\def\lres{\mathrel{%
		\mathchoice{\LRES}{\LRES}{\scriptstyle\LRES}{\scriptstyle\LRES}%
}}
\def\LRES{{%
		\setbox0\hbox{$\backslash$}%
		\rlap{\hbox to \wd0{\hss$\circ$\hss}}\box0
}}

\def\RRES{{%
		\setbox0\hbox{$/$}%
		\rlap{\hbox to \wd0{\hss$\circ$\hss}}\box0
}}

\def\checkmark{\tikz\fill[scale=0.4](0,.35) -- (.25,0) -- (1,.7) -- (.25,.15) -- cycle;} 
\numberwithin{thm}{section}
\usepackage{setspace}
\begin{document}
%\tableofcontents

\begin{frontmatter}
\newpage
\title{Max-algebraic hybrid automata: Modelling and equivalences}
%\tnoteref{mytitlenote}


\author[1]{Abhimanyu Gupta}
\ead{a.gupta-3@tudelft.nl}
\author[1]{Bart De Schutter}
\ead{b.deschuter@tudelft.nl}
\author[2]{Jacob van der Woude}
\ead{j.w.vanderwoude@tudelft.nl)}
\author[1]{Ton van den Boom}
\ead{a.j.j.vandenboom@tudelft.nl}
\address[1]{Delft Center for Systems and Control, Delft University of Technology, Delft, The Netherlands}
\address[2]{DIAM, EWI, Delft University of Technology, Delft, The Netherlands}

%\fnref{myfootnote}
%\tableofcontents
\begin{abstract}
This article introduces the novel framework of max-algebraic hybrid automata as a hybrid modelling language in the max-plus algebra. We show that the modelling framework unifies and extends the switching max-plus linear systems framework and is analogous to the discrete hybrid automata framework in conventional algebra. In addition, we show that the framework serves as a bridge between automata-theoretic models in max-plus algebra and switching max-plus linear systems. In doing so, we formalise the relationship between max-plus automata and switching max-plus linear systems in a behavioural sense. This also serves as another step towards importing tools for analysis and optimal control from conventional time-driven hybrid systems to discrete-event systems in max-plus algebra.

\end{abstract}

%\begin{keyword}
%\texttt{elsarticle.cls}\sep \LaTeX\sep Elsevier \sep template
%\MSC[2010] 00-01\sep  99-00
%\end{keyword}

\end{frontmatter}

\section{Introduction}
%{\color{blue}Please check again the parts in blue.}
%{\color{red}ROUTE 1: Complete as such by including an illustration. The result is that an SMPL system can capture both the external (input-output) and internal (allowed sequence of transitions) behaviour. For the illustration, figure out a simple example that can be written both as a max-plus automaton and an MAHA/SMPL system. Figure out how race policy and first-in-first-out policies are related. What can be do with a structural abstraction of a MAHA? How to deal with MIMO descriptions in the abstraction?}

%{\color{brown}ROUTE 2: FUTURE WORK Provide a procedure to construct (when possible) a bisimilar max-plus automaton from a given MAHA system. Requires a region-based finite abstraction. Max-plus, min-plus, and switching min-plus cases are known in literature. How can one go about (one-step reachability) an arbitrary max-min-plus-scaling system? Under what conditions can such a model lead to a max-plus automaton?}

Max-algebraic models are particularly suited for modelling discrete-event systems, with synchronisation but no concurrency or choice, when timing constraints on event occurrences are of explicit concern in system dynamics and performance specifications \cite{Baccelli1992,Cassandras2009,Komenda2018}. The modelling class coincides with that of timed-event graphs. Moreover, the modelling formalism provides a continuous-variable dynamic representation of discrete-event systems analogously to time-driven systems. This similarity has served as the key motivation in the development of max-plus linear systems theory, analogously to classical linear systems theory \cite{Baccelli1992,Cohen1999}. 

%Max-plus algebra plays a crucial role in studying the timing aspects of discrete-event systems governed by synchronisation but no concurrency \cite{Baccelli1992}. The resulting modelling formalism provides a continuous-variable dynamic representation of discrete-event systems analogously to time-driven systems. This similarity has served as the key motivation and driving force in the development of the max-plus linear systems theory analogously to the classical linear systems theory \cite{Cohen1999,Komenda2018}. 

The major limitation of the max-plus  linear modelling framework is rooted in its inability to model \textit{competition} and/or \textit{conflict} among several event occurrences \cite{Komenda2018}. The formalism then resorts to the dual min-plus operations to model conflict resolution policies explicitly in the algebraic system description \cite{Baccelli1992,Cohen1997a}. 

Automata-theoretic models for discrete-event systems, on the other hand, are particularly suited for modelling conflicts and certain forms of concurrency. To this end, models have been proposed in literature that follow a modular approach by allowing the conflict resolution mechanism be handled by a discrete variable taking values in a finite set \cite{Gaubert1995,VandenBoom2006}. The resulting hybrid phenomenon due to the interaction of the discrete-valued and continuous-valued dynamics is the \textit{focus of this article}. In this context, there are two layers of behaviour that are studied: logical ordering of the events on the one hand, and the timing of events on the other. 
%Different models, in general, play different roles and some are more suitable for certain applications than others.
  
The max-plus automata approach for modelling the aforementioned hybrid phenomenon forms an extension of finite automata where transitions are given weights in the max-plus algebra \cite{Gaubert1995}. The weight encodes the timing information as the price of taking the transition. The output under a given input sequence over an event alphabet is then evaluated, in the max-plus algebra, as the accumulated weight. Such models lend themselves to path-based performance analysis for discrete-event systems \cite{Gaubert1995,Gaubert1999}.  

An alternative approach involves the \Gls{SMPL} modelling paradigm \cite{VandenBoom2006}. Such models extend the max-plus linear modelling framework by allowing changes in the structure of synchronisation and ordering constraints as the system evolves \cite{VandenBoom2006}. This offers a compromise between the powerful description of hybrid systems and the decision-making capabilities in max-plus algebra \cite{DeSchutter2001,VandenBoom2006}. Moreover, the \Gls{SMPL} formalism offers the flexibility of explicitly modelling different switching mechanisms between the operating modes in a single framework \cite{VanDenBoom2012b}.

%{\color{red}However, this modelling formalism abstracts the mechanism by which transitions are orchestrated again into discrete events. This, in particular, allows modelling only the aggregated dynamics from one mode to the other. The conservativeness then lies in the difficulty in picking appropriate partitions of the state space for continuous or discrete control using mixed integer programs.}

%A controlled (non-autonomous) max-plus automata, where the supervisor is also modelled as a max-plus automata, has also also been studied as an extension to model disabling and/or modifying weights of transitions \cite{Komenda2008,Komenda2014}.

%In this article we focus on the \Gls{SMPL} modelling paradigm moving towards nonlinear expressions in max-plus algebra. Such models extend the max-plus linear modelling framework by allowing changes in the structure of synchronisation and ordering constraints as the system evolves over the event cycles \cite{VandenBoom2006}. This offers a compromise between the powerful description of hybrid systems and the decision-making capabilities in max-plus algebra \cite{DeSchutter2001,VandenBoom2006}. Moreover, the \Gls{SMPL} formalism offers the flexibility of explicitly modelling different switching mechanisms between the operating modes in a single framework \cite{VanDenBoom2012b}. Therefore, the \Gls{SMPL} modelling framework forms a subclass of timed Petri nets where the conflict resolution is carried out via a switching mechanism.  

In current article, we propose a novel \textit{max-algebraic hybrid automata} framework to model discrete-event systems analogously to the hybrid automata framework of \cite{Lygeros1998,Lygeros1999a} for conventional time-driven systems. In the proposed framework, the discrete-valued dynamics is represented as a labelled oriented graph and the continuous-valued dynamics is associated to each discrete state. We formally prove that this serves as a unifying framework for studying the aforementioned models and their equivalence relationships in the behavioural framework \cite{Polderman1998,Julius2005,VanderSchaft2004}. 
The paper is organised as follows. Section 2 gives some background on the
max-plus algebra. Section 3 reviews the literature on discrete-event systems in
max-plus algebra focusing on \Gls{SMPL} and max-plus automata frameworks.
Section 4 introduces the unifying modelling framework of max-algebraic hybrid
automata and its finite-state discrete abstraction. Section 5 establishes the
relationships among different modelling classes namely, \Gls{SMPL}, max-plus
automata, and the proposed max-algebraic hybrid automata. The paper ends with
concluding remarks in Section 6.
\iffalse
\section{Previous work}
\paragraph{Timed DES}: Classification

The event timing is important in assessing the performance of a DES often measured through
quantities such as throughput or response time. In these instances, we need to
consider the timed language model of the system. The fact that different event
processes are concurrent and often interdependent in complex ways presents
great challenges both for modeling and analysis of timed DESs.
\paragraph{Timed Petri nets} With and without multipliers; continuous/discrete/hybrid.
\paragraph{Timed automata}
\paragraph{(Max,+) Automata}
\paragraph{Max-min-plus algebra}
Supervisory control is the term established for describing the systematic means (i.e. enabling or disabling events which are controllable) by which the logical behaviour of a
DES is regulated to achieve a given specification.
\fi

\section{Preliminaries}\label{sec:Prem}
This section presents some basics in max-plus and automata theory based entirely on \cite{Baccelli1992,heidergott2014max,Olsder1991,Gunawardena1994,DeSchutter2004a,Cassandras2009}. 

The set of all positive integers up to $n$ is denoted as $\underline{n}=\{l\in\mathbb{N}\,|\,l\leq n\}$ where $\mathbb{N}=\{1,2,3,\dots\}$.

\textit{Max-plus algebra.} The max-plus semiring, $\mathbb{R}_{\mathrm{max}}=(\mathbb{R}_\varepsilon,\oplus,\otimes)$, consists of the set $\mathbb{R}_\varepsilon = \mathbb{R}\cup \{-\infty\}$ endowed with the addition $(a\oplus b = \mathrm{max}(a,b))$ and the multiplication $(a\otimes b = a+b)$ operations \cite{Baccelli1992}. The zero element is denoted as $\varepsilon = -\infty$ and the unit element as $\mathds{1}=0$. These elements are identities with respect to $\oplus$ and $\otimes$, respectively, and $\varepsilon$ is absorbing for $\otimes$. However, the max-plus algebra lacks an additive inverse operation (since $a\oplus b=\varepsilon$ implies $a=b=\varepsilon$). The matrix with all entries $\varepsilon$ is denoted as $\mathcal{E}$. The max-plus powers of a matrix $A\in{\mathbb{R}}_\varepsilon^{n\times n}$ are defined recursively as $\maxpow{A}{k+1} = \maxpow{A}{k}\otimes A$ for $k\in\mathbb{N}$. The partial order $\leq$ is defined such that for vectors $x,y\in\mathbb{R}^n_\varepsilon$, $x\leq y\Leftrightarrow x\oplus y = y \Leftrightarrow x_i\leq y_i,\; \forall i\in\underline{n}$. 

The max-plus vector and matrix operations can be defined analogously to the conventional algebra. Let $A,B\in\mathbb{R}^{m\times n}_\varepsilon$, and $C\in\mathbb{R}^{n\times p}_\varepsilon$; then
\begin{align*}
	[A \oplus B]_{i j}&=a_{i j} \oplus b_{i j}=\max \left(a_{i j}, b_{i j}\right) \\
	[A \otimes C]_{i j}&=\bigoplus_{k=1}^{n} a_{i k} \otimes c_{k j}=\max _{k}\left(a_{i k}+c_{k j}\right)
\end{align*}
where the $(i,j)$-th element of a matrix $A$ is denoted as $[A]_{ij}$ or $a_{ij}$. Likewise, the $i$-th element of a vector $x$ is denoted as $x_i$. 

The min-plus semiring, $\mathbb{R}_{\mathrm{min}}=(\mathbb{R}_\top,\oplus',\otimes')$ is defined as a dual of the max-plus semiring acting on the set $\mathbb{R}_\top=\mathbb{R}\cup \{+\infty\}$. The zero element is $\top = +\infty$. The vector and matrix operations are then defined analogously. The completed max-plus semiring is defined over the set $\overline{\mathbb{R}}_\varepsilon=\mathbb{R}_\varepsilon\cup \{\top\}$ such that max-plus operations take preference. The set of all vectors in $\overline{\mathbb{R}}^n_\varepsilon$ with at least one finite entry is denoted as $\overline{\mathbb{R}}^n_\varepsilon\setminus \{\varepsilon,\top\}^n$.

The max-plus Boolean semiring defined as $\mathbb{B}_{\max} = (\mathbb{B}_\varepsilon,\oplus,\otimes)$, where $\mathbb{B}_\varepsilon = \{\varepsilon,\mathds{1}\}$, is isomorphic to the Boolean semiring $\mathbb{B} = (\{\mathrm{false},\mathrm{true}\},\mathrm{or},\mathrm{and})$. 
%A max-plus zero matrix is denoted as $\mathcal{E}_{n\times n}$. A max-plus identity matrix $\mathcal{I}_{n}$ is defined as: $(\mathcal{I})_{ii} = 0$ for all $i\in\underline{n}$ and $(\mathcal{I})_{ij} = \varepsilon$ for all $i,j\in \underline{n}$ with $i\neq j$. The max-plus powers of a matrix are defined recursively as $\maxpow{A}{k+1} = \maxpow{A}{k}\otimes A$ for $k\in\mathbb{N}$ with $\maxpow{A}{0} = \mathcal{I}_{n}$. For scalars $\gamma,c\in\mathbb{R}$, we have $\maxpow{\gamma}{c}=c\cdot\gamma$. 
  %The partial order $\leq$ for vectors $x,y\in\mathbb{R}^n_\varepsilon$ is defined such that  $x\leq y\Leftrightarrow x\oplus y = y \Leftrightarrow x_i\leq y_i,\; \forall i\in\underline{n}$.  

%The min-plus semiring, $\mathbb{R}_{\mathrm{min}}=(\mathbb{R}_\top,\oplus',\otimes')$ is defined as a dual of the max-plus semiring acting on the set $\mathbb{R}_\top=\mathbb{R}\cup \{+\infty\}$ \cite{Baccelli1992}. The zero element is $\top = +\infty$. The vector and matrix operations are then defined analogously. The minimum operation $\oplus'$ is, however, nonlinear in the max-plus algebra. The interested reader is referred to \cite{Olsder1991,gunawardena2003max}, \cite[Section 9.6]{Baccelli1992} for an extensive survey on dynamical systems in max-min-plus algebra. The extended max-plus semiring $\overline{\mathbb{R}}_\varepsilon = \mathbb{R}_\varepsilon\cup\{+\infty\}$ is defined such that max-plus operations take preference. 

\textit{Max-min-plus-scaling functions.} The \Gls{MMPS} expression $f$ of the variables $x_1,\dots,x_n$ is defined by the grammar\footnote{The symbol $|$ stands for ``or". The definition is recursive.}
\begin{equation}\label{eq:0.1}
f:=x_i|\alpha|f_k\oplus f_l|f_k\oplus' f_l| f_k + f_l|\beta\cdot  f_k,\; \alpha,\beta\in \mathbb{R}, \; i\in \underline{n},
\end{equation}
where $f_k$ and $f_l$ are again MMPS expressions.

A max-min-plus expression $f$ of variables $x_1,\dots,x_n$ is defined by the
grammar
\begin{equation}
	f:=x_i|f_k\oplus f_l|f_k\oplus' f_l| f_k + \alpha,\; \alpha\in \mathbb{R}, \; i\in \underline{n},
\end{equation}
where $f_k$ and $f_l$ are again max-min-plus expressions. Any such max-min-plus expression $f$ can be placed in the max-min-plus conjunctive form:
\begin{equation}\label{eq:0.2}
	\begin{aligned}
		f &= f_1\oplus' f_2\oplus' \cdots \oplus' f_m,\\
		i&\neq j \Rightarrow f_i \nleq f_j,
	\end{aligned}
\end{equation}
where $f_j = (a_{j1}\otimes x_1)\oplus (a_{j2}\otimes x_2)\oplus \cdots\oplus
(a_{jn}\otimes x_n)$ is said to be a max-plus projection of $f$ with
$a_{ji}\in\mathbb{R}_\varepsilon$ for all $i\in\underline{n}$ and
$j\in\underline{m}$. The max-min-plus conjunctive form \eqref{eq:0.2} is unique
up to reordering of $f_j$'s \cite[Theorem 2.1]{Gunawardena1994}. Note that the
stated uniqueness is necessary for the definition of transition graphs (Definition \ref{def:1} below).
%{\color{blue}The stated uniqueness is necessary for the definition of
%transition graphs in Definition \ref{def:1}}

%There exists an $i\in\underline{n}$ such that $a_{ji}\neq \varepsilon$ for each $j\in\underline{m}$.

%\textit{Graph theory.} A directed graph $\mathcal{G}(A)= (V(A),E(A))$ can be associated to a matrix $A\in\mathbb{R}_\varepsilon^{n\times n}$ where the vertex set $V(A)=\underline{n}$ and the pair $(i,j)\in E(A)$, the edge set, whenever $a_{ji}\neq\varepsilon$.

%\textit{Function properties.} A function $g:\overline{\mathbb{R}}_\varepsilon^n \to\overline{\mathbb{R}}_\varepsilon^n$ is said to be \textit{additively homogeneous} if $g(\lambda + x) = \lambda + g(x)$, for all $\lambda\in\mathbb{R}$. The function $g$ is \textit{monotone} if for all $x,y\in\overline{\mathbb{R}}^n_\varepsilon$, $x\leq y$ implies $g(x)\leq g(y)$. 

\textit{Set theory.} Let $P$ be a finite set. Then $|P|$, $2^P$, and $P^*$ denote the cardinality, power set (set of all subsets), and set of non-empty finite sequences of elements from $P$, respectively. A non-empty finite set of symbols is referred to as an alphabet. When a set $P$ is a countable collection of variables, the set of valuations of its variables is denoted as $\mathbb{P}$. 

%\textit{Transition system.} A transition system is defined as a tuple $\mathcal{D} = (Q,\Sigma,\delta,Q_\mathrm{0},Y,H)$ consisting of a set of states $Q$, set of inputs $\Sigma$, a partial transition relation $\delta: Q\times \Sigma \to 2^Q$, a set of initial states $Q_\mathrm{0}\subseteq Q$, a set of outputs $Y$, and an output function $h:Q\to Y$. The trajectory of the system with length $\kappa\in\mathbb{N}\cup \{+\infty\}$ starting at $q_0\in Q_\mathrm{0}$ is a (possibly infinite) sequence $\left\{q_k,l_k,y_k\right\}_{k=1}^{\kappa}$ such that $q_k\in Q$, $l_k\in \Sigma$, $y_k \in Y$, $q_k \in \delta(q_{k-1},l_k)$ and $y_k = H(x_k)$ for all $k\in\underline{\kappa}$.  transition system, with an output in the Boolean semiring $\mathbb{B}$  

\textit{Finite automaton.} A finite automaton is a tuple $\mathcal{T}=(Q,\Sigma,\delta,Q_\mathrm{0},Q_\mathrm{f})$ consisting of a finite set of states $Q$, a finite alphabet of inputs $\Sigma$, a partial transition function $\delta : Q\times \Sigma\to 2^{Q}$, a non-empty set of initial states $Q_\mathrm{0}\subseteq Q$, and a non-empty set of final states $Q_\mathrm{f}\subseteq Q$.  A labelled transition is denoted as $q\xrightarrow[]{l} q'$ for $q'\in\delta(q,l)$, $l\in\Sigma$. 
%The output function, here, is replaced by an acceptance condition on the possible state trajectories.

A finite word is defined as a sequence (concatenation) of inputs $\omega_m = l_1 l_2\cdots l_m$. Here, $\omega_{m} = \omega_{m-1} l_{m}$, $l_{j}\in \Sigma$ for $j\in \underline{m}$. An empty word is denoted as $\epsilon$. An accepting path for a  word $\omega_m\in\Sigma^*$ on the finite automaton $\mathcal{T}$ is defined as the sequence of states $(q_0,q_1,\dots,q_m)\in Q^{m+1}$ if $q_0\in Q_\mathrm{0}$, $q_m\in Q_\mathrm{f}$, and $q_{i}\in \delta(q_{i-1},l_i)$ for all $i\in\underline{m}$. The set of words $\omega_m\in\Sigma^*$, $m\in\mathbb{N}$ accepted by some path on the finite automaton is denoted as $\llbracket \mathcal{T}\rrbracket_\mathrm{L}$, i.e. the language of the finite automaton $\mathcal{T}$. 

\section{Max-algebraic models of discrete-event systems}\label{sec:DES-model}
%The max-plus algebra provides a convenient approach for modelling discrete-event systems exhibiting synchronisation but no concurrency \cite{Baccelli1992}. This popularity and convenience is derived from the ability of this approach to represent the timing of discrete events as continuous valued linear dynamics. The automata approach, on the other hand, serves as a popular algebraic approach for modelling the logical aspects (choice and conflicts) of discrete-event systems \cite{Ramadge1989}. It is noted here that timed event graphs 
This section aims at recapitulating models in the max-plus algebra that capture synchronisation as well as certain forms of concurrency in discrete-event systems. For simplification of the exposition and for further systematic comparisons, we also present a common description of the underlying signals in discrete-event systems.  

\subsection{Synchronisation and concurrency}
%{\color{blue}I had earlier used the term \textit{parallelism} to describe the kind of concurrency discussed in this paper. I have since removed it to avoid confusion/complications.}
The max-algebraic modelling paradigm characterises the behaviour of a discrete-event system by capturing the sequences of occurrence times of events (or, \textit{temporal evolution}) over a discrete event counter. This, in particular, is useful when the events are ordered by the phenomena of synchronisation (max operation), competition (min operation), and time delay (plus operation) \cite{Baccelli1992}. The phenomenon of concurrency arising due to variable sequencing (and hence variable synchronisation and ordering structure) of events can lead to a semi-cyclic behaviour \cite{VandenBoom2020}. Below we discuss two different modelling approaches, namely \Gls{SMPL} systems and max-plus automata, that extend the max-plus linear framework to incorporate such concurrency. Here, the shared characteristic is the introduction of a discrete variable that completely specifies the ordering structure at a given event counter. This interaction of synchronisation and concurrency is, thus, \textit{hybrid} in nature.
\subsection{Signals in discrete-event systems}\label{sec:signal}
We refer to variables with finite or countable valuations as \textit{discrete}, and variables with valuations in $\overline{\mathbb{R}}_\varepsilon$ as \textit{continuous}. 
An event-driven system with both continuous and discrete variables evolving over a discrete counter $k$ is characterised by the following signals:
\begin{itemize}
	\item $x(\cdot)$ and  $l(\cdot)$: continuous and discrete states respectively;
	\item $u(\cdot)$ and $v(\cdot)$: continuous and discrete controlled inputs respectively;
	\item $y(\cdot)$: continuous output;
	\item $r(\cdot)$ and $w(\cdot)$: continuous and discrete exogenous inputs respectively. {The signal $r(\cdot)$ can represent a reference signal or a max-plus additive uncertainty in the continuous state $x$.} The signal $w(\cdot)$ can represent a scheduling signal or uncertainty in mode switching;
	\item $p(\cdot)$: continuous exogenous signal. {It can represent a parametric or max-plus multiplicative perturbation in the continuous dynamics that is either exogenous or state-dependent.}  
\end{itemize}	
The uncontrolled exogenous inputs, hereafter, are collected into a single signal $\Theta(\cdot)$. This signal is partitioned as $\Theta = [\Theta_\mathrm{x}^\top,\;\Theta_\ell^\top]^\top$. Here $\Theta_\mathrm{x} = [r^\top, \; p^\top]^\top$ denotes the uncertainty in the continuous-state evolution, and $\Theta_\ell = w$ denotes the uncertainty in the discrete-state evolution. 

%\subsection{\color{blue}Discrete abstraction of a max-plus linear system!}

\subsection{Switching max-plus linear systems}\label{sec:SMPL}
The dynamics of a general model in max-plus algebra in mode $l(k)\in\mathcal{L}\triangleq\underline{n_\mathrm{L}}$ for the continuous state $x(k)\in \overline{\mathbb{R}}^n_\varepsilon$ at event counter $k\in\mathbb{N}$ can be written as follows:
\begin{equation}\label{eq:31}
\begin{aligned}
x(k) &= f(l(k),x(k-1),u(k),\Theta_\mathrm{x}(k)), \\
l(k) &= \phi(l(k-1),x(k-1),  u(k),v(k),\Theta_\ell(k)), \\
y(k) &= h(l(k),x(k),u(k),\Theta_\mathrm{x}(k))
\end{aligned}
\end{equation}  
where the functions $f(\cdot)$ and $h(\cdot)$ represent the evolution of the continuous state and output, respectively, as \Gls{MMPS} functions. The function $\phi(\cdot)$ encodes the switching mechanism. 

We refer to an open-loop \Gls{SMPL} system, $\mathcal{S}_\mathrm{O}$, when the functions $f$ and $g$ are max-plus linear in states and inputs for a fixed $l$ and control inputs $u$ and $v$ are absent. On the other hand, we refer to a \textit{controlled} \Gls{SMPL} system, $\mathcal{S}_\mathrm{C}$, when a controller is also part of the system description. The control inputs in \eqref{eq:31} can then be modelled as outputs of a control algorithm:
\begin{equation}\label{eq:32}
\begin{aligned}
u(k) &= f^{(u)}_\mathrm{C}(z(k), \Theta(k))\\
v(k) &= f^{(v)}_\mathrm{C}(z(k), \Theta(k)).
\end{aligned}
\end{equation}
Here, the signal $z(\cdot)$ denotes the performance signal composed of the (past) known values of the continuous and discrete states, and continuous inputs. It is noted here that the functions $f^{(\cdot)}_\mathrm{C}$ might not have a closed form. The most popular control algorithms for continuous-valued discrete-event systems in literature are residuation \cite{Maia2003} and model predictive control \cite{VanDenBoom2012b}. 

An important subclass of controlled \Gls{SMPL} systems can be represented using
max-min-plus linear functions. This encompasses the class of max-plus linear
systems in open-loop and closed-loop with static \cite{VandenBoom2006} and
certain dynamic feedback controllers (for e.g., via residuation
\cite{Lahaye2008}). The max-min-plus linear functions can also be used to model
the dynamics of a subclass of timed Petri nets under a first-in first-out policy
\cite{Olsder1991,SotoyKoelemeijer2003}.

\begin{figure}[h]
	\centering
	\includegraphics[width=0.45\textwidth]{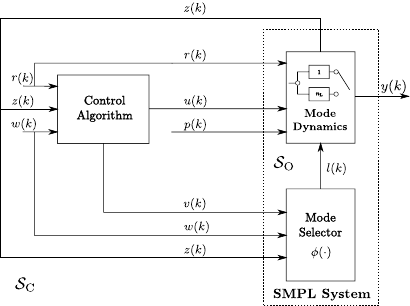}
	\caption{\label{fig:2} The \Gls{SMPL} system in closed loop $\mathcal{S}_\mathrm{C}$ with a controller, represented as a subclass of the class of discrete hybrid automata.}
\end{figure}

A controlled \Gls{SMPL} system can be represented as the connection of an \Gls{MMPS} dynamics $f$ in \eqref{eq:31} and a controller dynamics $f_\mathrm{C}$ in \eqref{eq:32} via a switching mechanism $\phi$ in \eqref{eq:31}. The controlled \Gls{SMPL} system can then be represented as a modification of discrete hybrid automata proposed in \cite{Torrisi2004} as shown in Fig.~\ref{fig:2}. 
%This is possible due to the equivalence of the \Gls{MMPS} and the piecewise affine dynamics under boundedness of the states and inputs of the system \cite{Heemels2001}. 
The major differences between our framework and that of \cite{Torrisi2004} are: \rn{1} the control algorithm is explicitly included in the model description, and \rn{2} the mode selector can also model a discrete dynamic process. The mode dynamics, however, is still piecewise affine due to the equivalence of max-min-plus-scaling and piecewise-affine systems under fairly non-restrictive assumptions on boundedness and well-posedness of the dynamics \cite{Heemels2001,VanDenBoom2012b}.   

In the sequel, we will adopt a more general representation for the transition notation. We denote by $(l^+,x^+)$ the successors of the current global state $(l,x)$. Similarly, we denote by $(l^-,x^-)$ the known state information that could possibly contain some parts of the current global state $(l,x)$ \cite{VanDenBoom2002a}. 

%Specific choices of the form of the functions $f$ and $g$ will determine different classes of hybrid systems
%\subsubsection{\color{blue}Uncertainty Model}
%{\color{blue}Descriptions of classes $\mathscr{R}$, $\mathscr{W}$ and $\mathscr{P}$.}
\subsubsection{Switching mechanism} \label{sec:switch}
The dynamic evolution of the discrete state $l$ can be brought about by either \rn{1} a discontinuous change in the continuous dynamics $f(\cdot)$ when the states satisfy certain constraints, or \rn{2} in a non-autonomous response to an exogenous event occurrence via the signal $w(\cdot)$. We refer to the dynamics as \textit{autonomous} in the absence of exogenous inputs $\Theta$.

We refer to the discrete evolution as \textit{controlled} when the controller (via \eqref{eq:32}) is incorporated into the system description in \eqref{eq:31}. 
%An additional structure can be imposed on the non-autonomous evolution when the switching sequence is constrained by, for e.g., a finite state machine.  
%\cbstart
Now we classify the switching mechanisms due to autonomous/non-autonomous and controlled/uncontrolled behaviour \cite{VandenBoom2006,VanDenBoom2012b}. The notions are then sub-classified in increasing order of complexity, where the first case(s) are special cases of the last one:
\begin{enumerate}[labelwidth=!, labelindent=0pt, label=\arabic*.]
	\item State-dependent switching: The function $\phi$ does not depend on exogenous inputs. The switching class can be segregated based on the presence or absence of controllers:
	\begin{enumerate}[label=\arabic{enumi}\alph*.]
		\item Autonomous switching: The controller is either absent or contains only memoryless maps that can be incorporated in the dynamics $f$:
		\begin{equation}
			\begin{bmatrix}
			(l^+)^\top &x^\top
			\end{bmatrix}^\top = f_\phi(l,x^-).
		\end{equation}
		\item Autonomous controlled switching: The control algorithm, in this case, is explicitly part of the system description:  
		\begin{equation}
			\begin{bmatrix}
			(l^+)^\top& x^\top&(u^+)^\top&(v^+)^\top
			\end{bmatrix}^\top = f_{\phi,\mathrm{C}}(l,x^-,u,v).
		\end{equation}
	\end{enumerate}
	\item Event-driven switching: The function $\phi$ depends only on discrete inputs (exogenous or controlled) and discrete states. The class can be subdivided as follows:
	\begin{enumerate}[label=\arabic{enumi}\alph*.]
		\item Externally driven switching: The switching sequence is completely (arbitrarily) specified by a discrete exogenous input. Therefore, the switching happens uncontrollably in response to exogenous events: 
		\begin{equation}
			l = \phi(\Theta_\ell).
		\end{equation}
		\item Constrained switching: The switching sequence is driven by a discrete exogenous input with constraints on allowed sequences:  
		\begin{equation}\label{eq:37}
		l = \phi(l^-,\Theta_\ell).
		\end{equation} 
		\item Constrained controlled switching: A combination of an exogenous discrete input and a discrete control input together describe the switching sequence along with constraints on allowed sequences:  
		\begin{equation}
		 l = \phi(l^-,v,\Theta_\ell).
		\end{equation}
	\end{enumerate}
\end{enumerate}  

\subsection{Max-plus automata}\label{sec:MPA}
The max-plus automata are a quantitative extension of finite automata combining the logical aspects from automata/language theory and timing aspects from max-plus linear system \cite{Gaubert1995}. Here, the concurrency is handled at the logical level of the finite automaton. The variable ordering structure in the sequence of events is brought about by the set of accepted input words. The transition labels are augmented with weights in the max-plus semiring. The continuous-variable output dynamics appears as a max-plus accumulation of these weights over the paths accepted by an input word. We now recall the formal definition to elucidate the functioning of a max-plus automaton. 
{\begin{figure}
	\centering
	\includegraphics[width=0.45\textwidth]{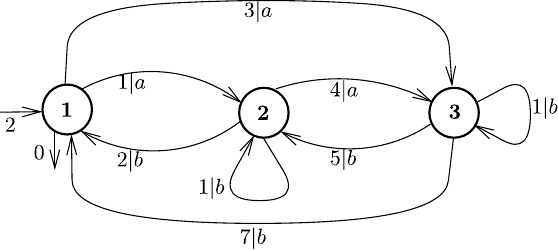}
	\caption{\label{fig:6} A nondeterministic max-plus automaton \cite{Gaubert1995}. An edge label consists of an event label (a letter in $\Sigma$) and a weight in the max-plus semiring $\mathbb{R}_{\max}$.}
\end{figure}}   
%{\color{red}CLARIFY!} {\color{blue}The parallelism, here, is directly handled by a finite state machine that allows to model different possible schedules/tasks of operations simultaneously.} The timing aspects of the events appear as a set of max-plus linear equations that are a function of the schedule thereby modelling the synchronisation phenomena. 
\begin{defn}[\cite{Gaubert1995}]\label{def:MPA}
%	\footnote{The max-plus automaton model in the discrete-event systems context was initially proposed in \cite{Gaubert1995}. It was then formulated in the standard automaton description by ignoring the initial and final weights in \cite{Komenda2009}.}
A max-plus automaton is a weighted finite automaton over the max-plus semiring $\mathbb{R}_{\mathrm{max}}$ and a finite alphabet of inputs $\Sigma$ represented by the tuple 
\begin{equation}\label{eq:39}
\mathcal{A} = (S,\alpha,\mu,\beta),
%\mathcal{A} = (S,\Sigma,\alpha,T,\beta),
\end{equation}
consisting of \rn{1} $S$, a finite set states, \rn{2} $\alpha:S\to \mathbb{R}_\varepsilon$, the initial weight function for entering a state, \rn{3} $\mu:\Sigma\to \mathbb{R}^{S\times S}_\varepsilon$, the transition weight function, and \rn{4} $\beta: S\to \mathbb{R}_{\color{blue}\varepsilon}$, the final weight function for leaving a state.\hfill $\blacklozenge$
%\begin{itemize}[labelwidth=!, labelindent=0pt]
%	\item 
%	\item 
	%$$, the initial time delay,

	%, the final time delay,
%	\item 
%		\item 
%	\item kl
	%It can also be represented as a set of transition time durations $T:Q\times \Sigma\times Q\to\mathbb{R}_\varepsilon$ such that $\mu(a)_{qq'}\triangleq T(q,a,q')$.    
%\end{itemize}
\end{defn}
%{\color{blue} I have kept the element $\varepsilon$ in the codomain of the weight functions $\alpha$ and $\beta$. The intuition is that a state $s\in S$ such that $\alpha(s)=\varepsilon$ can not be an initial state of the system. A similar argument holds for accepting states via $\beta(\cdot)$. It also helps in translation to vectors $\boldsymbol{\alpha}$ and $\boldsymbol{\beta}$ in \eqref{eq:11}.}

A labelled transition between $s,s'\in S$ is denoted as $s\xrightarrow[]{l\mid c} s'$ such that $[\mu(l)]_{ss'}=c$ for $l\in\Sigma$. The initial and final transitions are denoted as $\xrightarrow[]{c_\mathrm{0}} s$ and $s'\xrightarrow[]{c_\mathrm{f}}$ such that $\alpha(s) = c_\mathrm{0}$ and $\beta(s') = c_\mathrm{f}$, respectively. This can be represented by a weighted transition graph (Fig.~\ref{fig:6}).

The discrete (logical) evolution of a max-plus automaton for a given word $\omega_k = l_1l_2\cdots l_k\in\Sigma^*$ for $k\in\mathbb{N}$ is obtained by concatenating the labelled transitions as an accepting path $\rho_k=(s_0,s_1,\dots,s_k)\in S^{k+1}$ such that $\alpha(s_0)\neq \varepsilon$, $\beta(s_k)\neq \varepsilon$, and $[\mu(l_i)]_{s_{i-1}s_i}\neq \varepsilon$ for all $i\in\underline{k}$. The language of $\mathcal{A}$ is defined, analogously to that of a finite automaton, as the set of finite words accepted by the max-plus automaton: $\llbracket\mathcal{A}\rrbracket_\mathrm{L} = \{\omega_k\in\Sigma^*\mid \exists \rho_k\in S^* \;\text{s.t. } \rho_k \;\text{accepts } \omega_k \text{ with } k\in\mathbb{N} \}$. 

The continuous-valued trajectories of a max-plus automaton appear as the maximum
accumulated weight over all accepted discrete trajectories. Therefore, it can be
expressed completely using max-plus operations on the weights of the transition
labels. The output of the max-plus automaton $\mathcal{A}$ for the given word
$\omega_k$ is obtained over all accepting paths $\rho$ as
\begin{equation}\label{eq:map-io}
	\begin{aligned}
		y(\omega_k) := \underset{\rho\in S^{k+1}}{\max} \:&\left\{\alpha(s_0)\right. + [\mu(l_1)]_{s_{0}s_1} +[\mu(l_2)]_{s_{1}s_2} +\\ & \cdots + [\mu(l_k)]_{s_{k-1}s_k} + \left. \beta(s_k)\right\}. 
	\end{aligned}
\end{equation}
Given $n$ states in $S$, the initial weights
$\boldsymbol{\alpha}\in\mathbb{R}^{n}_\varepsilon$ and final weights
$\boldsymbol{\beta}\in\mathbb{R}_\varepsilon^n$ can be identified as vectors and
$\boldsymbol{\mu}(l)\in\mathbb{R}_\varepsilon^{n\times n}$ can be identified as
a matrix for all $l\in\Sigma$. Then the evolution of the continuous-valued
dynamics of the max-plus automaton $\mathcal{A}$ can be represented
as \cite{Gaubert1995}:
\begin{equation}\label{eq:11}
	\begin{aligned}
		x(\omega_{k}) &= x(\omega_{k-1})\otimes \boldsymbol{\mu}(l_k),\quad  x(\epsilon) = \boldsymbol{\alpha}^\top\\
		y(\omega_k)&=\boldsymbol{\alpha}^\top \otimes \boldsymbol{\mu}(l_1)\otimes\boldsymbol{\mu}(l_2)\otimes\cdots\otimes\boldsymbol{\mu}(l_k)\otimes \boldsymbol{\beta}\\
		%&=\boldsymbol{\alpha}^\top \otimes \boldsymbol{\mu}(\omega_k)\otimes \boldsymbol{\beta}\\
		&=x(\omega_k)\otimes \boldsymbol{\beta}.
	\end{aligned}
\end{equation}
%{\color{blue}Please refer to the discussion on the usage of the phrase finite-state abstraction at the beginning of Section 4.2.}

The finite-state discrete abstraction of a max-plus automaton is a finite automaton. It can be obtained by restricting the weights on transitions of $\mathcal{A}$ to the Boolean semiring $\mathbb{B}$ \cite{Gaubert1995}: 
\begin{equation}\label{eq:13}
\mathcal{A}_\mathrm{T} = (S,\Sigma,\delta_\mathcal{A},S_\mathrm{0},S_\mathrm{f}),
\end{equation}
where the partial transition relation $\delta_\mathcal{A}:S\times \Sigma\to 2^S$
is defined such that $s'\in \delta_\mathcal{A}(s,l)$ if $[\mu(l)]_{ss'}\neq
\varepsilon$. Similarly, we have $s\in S_\mathrm{0}$ if $\alpha(s)\neq
\varepsilon$ and $s'\in S_\mathrm{f}$ if $\beta(s')\neq \varepsilon$. The
acceptance condition for a word by the automaton $\mathcal{A}_\mathrm{T}$
follows immediately \cite{Gaubert1995}. Moreover, the max-plus automaton and its
finite-state discrete abstraction share the same language, i.e. $\llbracket
\mathcal{A}_\mathrm{T}\rrbracket_\mathrm{L}=\llbracket
\mathcal{A}\rrbracket_\mathrm{L}$. 

\begin{exmp}\label{eg:MPA}
	A max-plus automaton (from \cite{Gaubert1995}) with states $S = \{1,2,3\}$ over finite alphabet $\Sigma = \{a,b\}$ is depicted in Fig.~\ref{fig:6}. The transition weight functions can be represented as matrices of appropriate dimensions:
	\begin{equation}\label{eq:8}
			\begin{aligned}
				\boldsymbol{\mu}(a) &= \begin{pmatrix}
					{\varepsilon} & {1} & {3} \\
					{\varepsilon} & {\varepsilon} & {4} \\
					{\varepsilon} & {\varepsilon} & {\varepsilon}
					\end{pmatrix},
				&\boldsymbol{\mu}(b) &= \begin{pmatrix}
						{\varepsilon} & {\varepsilon} & {\varepsilon} \\
						{2} & {1} & {\varepsilon} \\
						{7} & {5} & {1}
						\end{pmatrix}\\
				\boldsymbol{\alpha}&= \begin{pmatrix}
							{0} &{\varepsilon} & {\varepsilon}
							\end{pmatrix}^\top,
				&\boldsymbol{\beta} &= \begin{pmatrix}
						{2} & {\varepsilon} & {\varepsilon}
						\end{pmatrix}^\top.
			\end{aligned} 
	\end{equation}
	The generated language can be obtained from the event labels of the paths originating from the initial state $1$ and terminating at the final state $2$ in Fig.~\ref{fig:6}. Such words $\omega\in\Sigma^*$ are of the form $ab$, $aab$, $aabb$, and so on.
\end{exmp}

We can now proceed to the introduction of a unified modelling framework represented by a max-algebraic hybrid dynamical system.
%Finally, we would like to point out that similarity between the \Gls{SMPL} and max-plus automata frameworks \cite{Gaubert1995}.  
%%%%%%%%%%%%%%%%%%%%%%%%%%%%%%%%%%%%%%%%%%%%%%%%%%%%%%%%%%%%%%%%%%%%%%%%%%%%%%%%%%%%%%%%%%%%%%%%%%%%%%%%%%%%%%%%%%%%%%%%%%%%%%%%%%%%%%%%%%%%%%%%%%%%%%
\iffalse 

The nondeterminism in max-plus automata can be best understood from its underlying conventional automata. Let $\delta:Q\times \Sigma\to 2^Q$ denote the underlying (partial) transition map of the max-plus automata, and let $Q_i$ denote the set of initial states:
\begin{equation}
\begin{aligned}
\delta(q,a) &= \left\{q'\mid \mu(a)_{qq'}\neq \varepsilon\right\},\\
Q_i &= \left\{q\mid \alpha(q)\neq \varepsilon\right\}.
\end{aligned}
\end{equation}
Then the max-plus automaton is said to be \textit{deterministic} if the set of initial states is singleton ($Q_i = \{q_0\}$) and for all $q\in Q$ and $a\in \Sigma$, $\delta(q,a)$ is also singleton. This implies that there is at most one finite entry in $\alpha$ and in each row of $\mu(a)$ for all $a\in \Sigma$. Else, the automaton is said to be nondeterministic as depicted in Fig.~\ref{fig:2}. This is referred to as \textit{structural nondeterminism} in the timed discrete-event system framework of \cite{Brandin1994}. However, there is \textit{no} nondeterminism in the timing aspects of the transitions in max-plus automata.  
\fi
%%%%%%%%%%%%%%%%%%%%%%%%%%%%%%%%%%%%%%%%%%%%%%%%%%%%%%%%%%%%%%%%%%%%%%%%%%%%%%%%%%%%%%%%%%%%%%%%%%%%%%%%%%%%%%%%%%%%%%%%%%%%%%%%%%%%%%%%%%%%%%%%%%%%%%
%\cbend 
\section{Unified modelling framework}\label{sec:DEHA}
We propose a novel modelling framework of \textit{max-algebraic hybrid automata}
for discrete-event systems as hybrid dynamical systems in the max-plus algebra
\eqref{eq:31}. The modelling language  allows composition with
controllers/supervisors and abstraction to refine design problems for individual
components. We also propose a finite-state discrete abstraction of the
max-algebraic hybrid automaton that preserves the allowed ordering of events of
the discrete-event system. 
%{\color{blue} I had referred to the untimed aspects of ordering of events of a discrete-event system as its `logical' property/modelling. I have since removed the term.}   

Later, we show (in Section \ref{sec:Relation}) that the proposed max-algebraic hybrid automata framework also serves as a link between \Gls{SMPL} systems and max-plus automata. The proposed modelling framework is more descriptive than \Gls{SMPL} systems and max-plus automata in that it allows to capture the different types of interactions between the continuous and discrete evolutions (as presented in Section \ref{sec:switch}). Most importantly, the model retains the structure of switching between dynamical systems.   
%We show that the proposed unified modelling framework can model the timing aspects of event graphs (parallel evolution on a set of continuous states) along with the logical aspects of ordering of events via mode switching (sequential evolution on a set of discrete states).
%The continuous state evolves according to a set of max-plus linear system of equations. The discrete state evolution is left to a mode selector based on an exogenous discrete signal, a controlled discrete input, and a logic signal from the continuous part. The discrete state triggers switching of continuous dynamics (among predefined modes). This interaction of the continuous and discrete dynamics results in a max-algebraic hybrid automata. When the inputs to the hybrid system are derived from a controller, we refer to a controlled max-algebraic hybrid system model. 
%\cbstart
%To this end, we propose a controlled max-algebraic hybrid system modelling framework as an extension of hybrid automata framework in \cite{Kowalewski2011}.

\subsection{Max-algebraic hybrid automata}\label{sec:MAHA}
A max-algebraic hybrid automaton is presented as an extension of the open hybrid automata in \cite{Lygeros1998,Lygeros1999} to incorporate max-algebraic dynamics. 
%{\color{red}Our model is different from \cite{Reveliotis2017} in that \dots}
\begin{defn}\label{def:MAHA}
	A max-algebraic hybrid automaton with both continuous and discrete inputs and distinct operating modes can be represented as a tuple
	\begin{equation}\label{eq:41}
	\mathcal{H} = \left( Q,\mathbb{X},\mathbb{U},\mathbb{V},\mathbb{Y},\mathrm{Init},F,H,\mathrm{Inv},E,G,R,\Lambda\right)
	\end{equation}
	where:
	\begin{itemize}[labelwidth=!, labelindent=0pt]
		\item $\mathbb{Q}$ is a finite set of discrete states (or, modes);
		%   \item $X:=\{X_q\}_{q\in Q}$, $X_q\subseteq \overline{\mathbb{R}}_\varepsilon^{n}$ is a set of continuous states.
		%	\item $U:=\{U_q\}_{q\in Q}$, $U_q\subseteq \overline{\mathbb{R}}_\varepsilon^{n_\mathrm{u}}$ is a set of continuous input variables.
		%		\item $V:=\bigcup_{q\in Q}V_q$, $V_q\subseteq {\underline{n_\mathrm{L}}}$ is a set of discrete input variables.
		%	\item $V:=\{V_q\}_{q\in Q}$, $V_q\subseteq {\underline{n_\mathrm{L}}}$ is a set of discrete input variables.
		%	\item $Y:=\{Y_q\}_{q\in Q}$, $Y_q\subseteq \overline{\mathbb{R}}_\varepsilon^{n_\mathrm{y}}$ is a set of continuous output variables.	
		%\item $\mathbb{X}:=\bigcup_{q\in Q}\mathbb{X}_q$, $\mathbb{X}_q\subseteq \overline{\mathbb{R}}_\varepsilon^{n}$ is a set of continuous states.
		\item $\mathbb{X}\subseteq \overline{\mathbb{R}}_\varepsilon^{n}$ is the set of continuous states;
		\item $\mathbb{U}\subseteq \overline{\mathbb{R}}_\varepsilon^{n_\mathrm{u}}$ is the set of continuous inputs;
		\item $\mathbb{V}$ is a finite set of discrete inputs;
		\item $\mathbb{Y}\subseteq \overline{\mathbb{R}}_\varepsilon^{n_\mathrm{y}}$ is the set of continuous outputs;
		\item $\mathrm{Init}\subseteq \mathbb{Q}\times \mathbb{X}$ is the set of initial states;
		\item $F:\mathbb{Q}\times\mathbb{X}\times  \mathbb{U} \to \mathbb{X}$ is the continuous-valued dynamics associated to each mode $q\in \mathbb{Q}$;
		\item $H:\mathbb{Q}\times\mathbb{X}\times \mathbb{U}\to \mathbb{Y}$ is the continuous-valued output equation;
		\item $\mathrm{Inv}: \mathbb{Q}\to 2^{\mathbb{X}\times \mathbb{U}\times \mathbb{V}}$ assigns to each $q\in  \mathbb{Q}$ an invariant domain specifying a set of admissible valuations of the state and input variables.
%		\item $E\subseteq  \mathbb{Q}\times  \mathbb{Q}$ is the set of possible discrete transitions and its elements are defined as $\eta = (q,q')$, where $q$ is the starting mode and $q'$ is the terminal mode of the edge;
		%\item $G^\mathrm{A}:=\{G^\mathrm{A}_q\}_{q\in Q}$, $G^\mathrm{A}_q\subset \mathrm{Inv}_q$, is the collection of autonomous guard sets which specify when autonomous switching from a particular mode is possible.
		%\item $\delta^\mathrm{A}:=\{\delta^\mathrm{A}_e\}_{e\in E}$, $  \delta^\mathrm{A}_e:G^\mathrm{A}_q\to X\times Q$, is the collection of autonomous switching maps between the modes. 
		\item $G:E\to 2^{\mathbb{X}\times\mathbb{U}\times\mathbb{V}}$ is the collection of guard sets which assigns to each edge $\eta=(q,q')\in E$ the admissible valuation of the state and input variables when transition from the mode $q$ to $q'$ is possible; 
		%\item $V:\{V_q\}_{q\in Q}$, is the collection of mode transition control sets. Here, $n_q$ $(\leq n_\mathrm{L})$ denotes the number of all possible mode transitions from a mode $q$. These can be modelled as Boolean signals for forced (controlled or uncontrolled) mode switching.
		%\item $\delta:\, G\times V \to X\times Q$ is the discrete transition/switching map from a given mode onto a subset of the max-algebraic hybrid state space $S=X\times Q$. 
		%	\item $\delta:=\{\delta_q\}_{q\in Q}$, $  \delta_q:G_q\times V_q\to X\times Q$, is the collection of transition maps between the modes parametrised by a subset of transition control set $V$. This map gives a subsequent mode of operation and also the intersection specification of the continuous states from one mode to another.  
		%	\item $G^\mathrm{C}:=\{G^\mathrm{C}_q\}_{q\in Q}$, $G^\mathrm{C}_q\subset \mathrm{Inv}_q$, is the collection of controlled guard sets which specify when controlled switching from a particular mode is possible. 
		%	\item $\delta^\mathrm{C}:=\{\delta^\mathrm{C}_e\}_{e\in E}$, $  \delta^\mathrm{C}_e:G^\mathrm{C}_q\times V_q\to X\times Q$, is the collection of controlled switching maps between the modes parametrised by a subset of transition control set $V$.      
		\item $R:=E\times \mathbb{U}\times\mathbb{V}\to 2^{\mathbb{X}\times\mathbb{X}}
		$ is the collection of reset maps which assigns to each edge $\eta=(q,q')\in E$, $u\in\mathbb{U}$ and $v\in \mathbb{V}$ a destination map specifying the continuous states before and after a discrete transition; 
		\item $\Lambda: \mathbb{Q}\times \mathbb{X}\to 2^{\mathbb{U}\times \mathbb{V}}$ assigns to each state a set of admissible inputs. \hfill $\blacklozenge$
	\end{itemize}
\end{defn}
The hybrid state of the max-algebraic hybrid automaton $\mathcal{H}$ is given as $(q,x)\in \mathbb{Q}\times\mathbb{X}$.
%It is noted that any possible evolution of the hybrid state $\mathbb{H}$ described by the max-algebraic hybrid automaton $\mathcal{H}$ is discrete along a counter $k\in\mathbb{N}$. 
The hybrid nature stems from the interaction of the discrete-valued state $q\in \mathbb{Q}$ and the continuous-valued state $x\in\mathbb{X}$. Moreover, the valuations of the continuous variables of $\mathcal{H}$ are defined over the completed max-plus semiring $\overline{\mathbb{R}}_\varepsilon$. Therefore, the proposed max-algebraic hybrid automaton forms a novel extension of the hybrid automata framework in \cite{Lygeros1999a}.   

The hybrid state of the max-algebraic hybrid automaton $\mathcal{H}$ is subject to change starting from $(q_0,x_0)\in\mathrm{Init}$ as concatenations of \rn{1} discrete transitions in the continuous-valued state, to $(q_0,x)$, according to $x = F(q_0,x_0,\cdot)$, as long as the invariant condition of the mode $q_0$ is satisfied, i.e. $(x,\cdot,\cdot)\in\mathrm{Inv}(q_0)$, and \rn{2} discrete transitions in the mode, $(q_0,x)$ to $(q',x')$, as allowed by the guard set $(x,\cdot,\cdot)\in G(\eta)$, $\eta = (q_0,q')$, while the continuous-valued state changes according to the reset map $(x,x')\in R(\eta,\cdot,\cdot)$. 
%A transition is said to be \textit{instantaneous} if it does not incur a change in the event counter $k$.

The exogenous inputs $u\in \mathbb{U}$ and $v\in \mathbb{V}$ allowed by a given hybrid state $(q,x)$, or $(u,v)\in \Lambda(q,x)$, can affect the system evolution through: \rn{1} the continuous-valued mode dynamics $x' = F(q,x,u)$ and $y = H(q,x,u)$ when $(x,u,v)\in \mathrm{Inv}(q)$, \rn{2} the guard sets $(x',u,v)\in G(\eta)$ allowing discrete mode transitions along $\eta=(q,q')\in E$, \rn{3} the mode invariants $(x',u,v)\notin\mathrm{Inv}(q)$ forcing discrete mode transitions, and \rn{4} the reset maps $(x',x'')\in R(\eta,u,v)$.

%Given a sequence of event counters $k\in\mathbb{N}$ A possible evolution of the max-algebraic hybrid automaton $\mathcal{H}$ is defined over a discrete-event counter $k\in\mathbb{N}$ as follows: \rn{1} the hybrid state evolves according to $x^+ = f(q_0,x,u)$ from a given initial state $(q_0,x_0)\in \mathrm{Init}$ 
%The max-algebraic hybrid automata framework differs from the hybrid automata framework of \cite{Lygeros1996} in the following aspects:{ \color{red} No hybrid time, all dynamics is discrete.}. Firstly, the state, input and output variables are contained in the max-min-plus algebra. {\color{blue}Secondly, the model allows multiple discrete transitions in $E$ with the same guard and reset labels originating from a discrete state. This implies that the discrete state transition map is allowed to be set-valued in accordance with \cite{Cassandras2009}.}                                                                                     
%{\color{blue} As soon as the invariant condition is violated, the system is forced to leave the current location by taking the discrete transition. The discrete dynamics is restricted using guards. A discrete transition is enabled if the guard condition is true, i.e. valuation of the continuous variables fulfil the guard. The continuous variables are reinitialised during a discrete transitions according to the reset mapping. Whenever both invariant and guard conditions are simultaneously true, there is a choice between continuous and discrete evolution.}

\subsection{Finite-state discrete abstraction of max-algebraic hybrid automata}\label{sec:HT}
% %{\color{red}Read how Adzkiya solves the problem of multiple inputs and outputs \cite{Adzkiya2013,Adzkiya2014thesis}.}
% {\color{blue}I had used the term \textit{(purely) discrete abstraction} to
% stress that the continuous-valued components are abstracted away from the model.
% Discrete abstraction has now been replaced with finite-state abstraction
% everywhere in the paper. The problem is that max-plus automata are already
% finite-state systems but with continuous valued weights on the transitions. Then
% the term \textit{finite-state abstraction} in the context of max-plus automata
% in \eqref{eq:13} does not sound appropriate.}

We now propose a finite-state discrete abstraction of a max-algebraic hybrid automaton \eqref{eq:41} by embedding it into a finite automaton. The proposed discrete abstraction of $\mathcal{H}$ is a one-step transition system abstracting away valuations of the continuous variables while preserving the state-transition structure of the underlying discrete-event system. 

To this end, we define one-step state transition relations corresponding to the mode dynamics $F$ and $H$ based on the underlying directed graph. 

%\begin{assum}\label{assum:1}
%     It is assumed that there is no continuous-valued input $u\in\mathbb{U}$ to the system.
%\end{assum}

% \begin{assum}\label{assum:1}
%     It is assumed that there is no continuous-valued input $u\in\mathbb{U}$ to the system.
% \end{assum}
% The preceding assumption is not restrictive for studying the logical behaviour of the system. The different input signal values can still be modelled as a state of the system under different closed-loop structures \cite{Schullerus2003a}.  
%{\color{blue}The following assumption is not part of the automaton description in \eqref{eq:41}. The dynamics can, in general, be defined using max-min-plus-scaling functions. This is important for translation of closed-loop \Gls{SMPL} system into a max-algebraic hybrid automaton (Theorem \ref{thm:SC-H}). I have shortened the discussion on the assumption. The discussion on max-min-plus functions is moved to Section \ref{sec:SMPL}. The assumption on the reset map is now made before Proposition \ref{prop:HT} to obviate a discussion on its `discretisation'.}
\begin{assum}\label{assum:2}
	The dynamics $F:(q,x,u)\to F(x,q,u)$ and the output function $H:(q,x,u)\to
	H(x,q,u)$ in \eqref{eq:41} are max-min-plus functions of the state
	$x\in\mathbb{X}$ and the input $u\in\mathbb{U}$ for every mode $q\in
	\mathbb{Q}$.  Also, the reset relation is defined such that for a discrete
	transition allowed by the guard set (i.e. $(x,u,w)\in G(\eta)$ for $\eta =
	(q,q')$), we have $x = x'$ if $(x,x')\in R(\eta,u,w)$. We denote such a
	map as $R(\cdot):=R_\mathrm{id}(\cdot)$. 
\end{assum}
 The subclass of max-algebraic hybrid automata modelled using max-min-plus functions is large enough to characterise a broad range of discrete-event systems (see Section \ref{sec:SMPL}). The assumption on the reset relation signifies that the exogenous discrete input via $\mathbb{V}$ does not directly impact the continuous-valued state $x\in\mathbb{X}$.

%The class of max-min-plus linear systems encompasses the class of max-plus linear  systems in open-loop and closed-loop with static and certain dynamic feedback controllers (for e.g., via residuation \cite{Lahaye2008}). The max-min-plus linear functions can also be used to model the dynamics of a subclass of timed Petri nets under first-in first-out policy \cite{Olsder1991,SotoyKoelemeijer2003}. As the dynamics is discrete, the effect of continuous-valued input on the output can usually be modelled via the state update \cite{Baccelli1992}. 

%The class of continuous monotone and additively homogeneous functions includes the class of max-plus and max-min-plus functions \citep{Gunawardena1994}.  
For convenience, it is also assumed that the functions $F$ and $H$ are in the max-min-plus conjunctive form \eqref{eq:0.2}. The ambiguity resulting from unspecified ordering of the max-plus projections, in \eqref{eq:0.2}, is not of consequence to the following analysis.
%The restriction on the dynamics due to Assumption \ref{assum:2} allows obtaining the logical behaviour of the discrete-event system from the underlying directed graph of the continuous-valued state dynamics. 
Then, there exist $L,M\in\mathbb{N}$ such that the mode dynamics can be expressed as \cite{Gunawardena1994}: 
\begin{equation}\label{eq:4}
	\begin{aligned}
		x^+ &= F(q,x,u) &&= \min_{l\in \underline{L}} \left(A^{(q,l)}\otimes x\oplus B^{(q,l)}\otimes u\right),\\
		y &= H(q,x,u) &&= \min_{m\in \underline{M}} \left(C^{(q,m)}\otimes x\oplus D^{(q,m)}\otimes u\right).
	\end{aligned}
\end{equation}
Here, $A^{(q,l)}\in\overline{\mathbb{R}}_\varepsilon^{n\times n}$, $B^{(q,l)}\in\overline{\mathbb{R}}_\varepsilon^{n\times n_\mathrm{u}}$ and $C^{(q,m)}\in\overline{\mathbb{R}}_\varepsilon^{n_\mathrm{y}\times n}$ for all $q\in\mathbb{Q}$, $l\in\underline{L}$ and $m\in\underline{M}$.
%Let $e^{(n)}_{\{j\}}$ denote the $j$-th vector of the canonical basis of $\mathbb{R}^n$ such that $[e^{(n)}_{\{j\}}]_j = 1$ and $[e^{(n)}_{\{j\}}]_i = 0$ for all $i\in\underline{n}\setminus \{j\}$. 

We associate the sets of labels $X_{\mathrm{var}}=\{\mathbf{x}_1,\mathbf{x}_2,\dots,\mathbf{x}_n\}$, $U_{\mathrm{var}}=\{\mathbf{u}_1,\mathbf{u}_2,\dots,\mathbf{u}_{n_\mathrm{u}}\}$ and $Y_{\mathrm{var}} = \{\mathbf{y}_1,\mathbf{y}_2,\dots,\mathbf{y}_{n_\mathrm{y}}\}$ with the continuous-valued state, input and output variables, respectively. 
%The $j$-th unit vector of the identity matrix of size $c$ is denoted as $e^{(c)}_{\{j\}}$ such that $[e^{(c)}_{\{j\}}]_j = 1$ and $[e^{(c)}_{\{j\}}]_i = 0$ for all $i\in\underline{c}\setminus \{j\}$.  
\begin{defn}\label{def:1}
	Given that Assumption \ref{assum:2} is satisfied, the one-step state transition graph $\Gamma_{F}^{(q)}\subseteq (X_{\mathrm{var}}\times X_{\mathrm{var}})\cup (U_{\mathrm{var}}\times X_{\mathrm{var}})$ of the continuous dynamics $F(q,\cdot,\cdot)$, $q\in \mathbb{Q}$, is defined such that for $(i,j)\in \underline{n}^2$ and $(p,j)\in\underline{n_\mathrm{u}}\times\underline{n}$:
	\begin{equation}\label{eq:14}
		\begin{aligned}
		&(\mathbf{x}_i,\mathbf{x}_j)\in \Gamma_{F}^{(q)} \Leftrightarrow \{\exists l\in \underline{L}\; \mathrm{ s.t. }\; [A^{(q,l)}]_{ji} \;\mathrm{ is}\,\mathrm{finite}\},
		\\
		&(\mathbf{u}_p,\mathbf{x}_j)\in \Gamma_{F}^{(q)} \Leftrightarrow \{\exists l\in \underline{L}\; \mathrm{ s.t. }\; [B^{(q,l)}]_{jp} \;\mathrm{ is}\,\mathrm{finite}\}.
		\end{aligned}
	\end{equation}
	The one-step state transition graph $\Gamma_{H}^{(q)}\subseteq (X_{\mathrm{var}}\times Y_{\mathrm{var}})\cup (U_{\mathrm{var}}\times Y_{\mathrm{var}})$ of $H(q,\cdot,\cdot)$, $q\in \mathbb{Q}$, is defined such that for $(i,j)\in \underline{n}\times\underline{n_\mathrm{y}}$ and $(p,j)\in\underline{n_\mathrm{u}}\times\underline{n_\mathrm{y}}$:
	\begin{equation}\label{eq:16}
		\begin{aligned}
			&(\mathbf{x}_i,\mathbf{y}_j)\in \Gamma_{H}^{(q)} \Leftrightarrow \{\exists m\in \underline{M}\; \mathrm{ s.t. }\; [C^{(q,m)}]_{ji} \;\mathrm{ is}\,\mathrm{finite}\}\\
			& (\mathbf{u}_p,\mathbf{y}_j)\in \Gamma_{H}^{(q)} \Leftrightarrow \{\exists m\in \underline{M}\; \mathrm{ s.t. }\; [D^{(q,m)}]_{jp} \;\mathrm{ is}\,\mathrm{finite}\}.
		\end{aligned}
	\end{equation}
\end{defn}
The transition graph $\Gamma_F^{(\cdot)}$ corresponds to the support of the
dynamics $F$ in that the membership of a pair $(\mathbf{x}_i,\mathbf{x}_j)$ in
$\Gamma_\mathrm{F}^{(\cdot)}$ indicates whether the component $F_j$ is an
unbounded function of the coordinate $x_i$ or not. Similarly, the transition graph
$\Gamma_H^{(\cdot)}$ corresponds to the support of the output equation $H$.
\hfill $\blacklozenge$ 

%For a given finite alphabet $\mathbb{V}$ in \eqref{eq:41}, we define a labelling function $\phi:E\to \mathbb{V}$. 
% We also define a map $\mathcal{R}:E\times \mathbb{V} \to 2^{X_\mathrm{var}\times X_\mathrm{var}}$ as a finite-state version of the reset map such that for $\eta = (q,q')$ and $w\in\mathbb{V}$ we have
% \begin{equation}\label{eq:19}
% 	(\mathbf{x}_i,\mathbf{x}_j)\in \mathcal{R}(\eta,w)\Leftrightarrow \{(x,x')\begin{aligned}[t]&\in R(\eta,w) \\&\Rightarrow x_i,x_j'\;\mathrm{are}\,\mathrm{finite}\}.\end{aligned} 
% \end{equation} 
% {\color{blue}The preceding reset relation corresponds to the re-initialisation of the state variable in $X_\mathrm{var}$, with change of mode in $\mathbb{Q}$, in response to the exogenous input $w\in\mathbb{V}$.  
% }

We now propose a finite-state discrete abstraction of the max-algebraic hybrid
automaton \eqref{eq:41}. The mode dynamics of the max-algebraic hybrid automaton
is abstracted as a one-step transition system. Here, the one-step transition naturally
corresponds to the evolution of the discrete-event system in one event step
$k\in\mathbb{N}$. Therefore, we denote it with a unique label $1$.       
\begin{prop}\label{prop:HT}
    Consider a max-algebraic hybrid automaton $\mathcal{H}$ (as in \eqref{eq:41}) under Assumption \ref{assum:2}. We assume that\footnote{The set of all vectors in $\overline{\mathbb{R}}^n_\varepsilon$ with at least one finite entry is denoted as $\overline{\mathbb{R}}^n_\varepsilon\setminus \{\varepsilon,\top\}^n$.}, $\mathbb{X} = \overline{\mathbb{R}}^n_\varepsilon\setminus \{\varepsilon,\top\}^n$. Then the max-algebraic hybrid automaton $\mathcal{H}$ generates a finite automaton.
\end{prop}
    
\begin{pf}
	A finite automaton embedding a max-algebraic hybrid automaton can be generated as a one-step transition system:
\begin{equation}\label{eq:17}
            \mathcal{H}_\mathrm{T} = (\overline{Q},\overline{\Sigma},\delta_\mathcal{H},\overline{Q}_\mathrm{0},\overline{Q}_\mathrm{f}),
        \end{equation}
        that consists of: 
            \begin{itemize}
            \item the finite set of states $\overline{Q} = \mathbb{Q}\times (X_\mathrm{var}\cup U_\mathrm{var})$;
            \item the input alphabet as a union of mode transition event labels and the one-step transition label denoting state transitions within a mode, $\overline{\Sigma} = \mathbb{V} \cup \{1\}$;
            \item the set of initial states $\overline{Q}_\mathrm{0}$ with $(q,\mathbf{x}_j)\in \overline{Q}_\mathrm{0}\subseteq \mathbb{Q}\times (X_\mathrm{var}\cup U_\mathrm{var})$ if $(q,x)\in \mathrm{Init}$ and $x_j \neq \varepsilon$, and $(q,\mathbf{u}_p)\in \overline{Q}_\mathrm{0}$ if there exists $\mathbf{x}_j\in X_\mathrm{var}$ such that $(\mathbf{u}_p,\mathbf{x}_j)\in\Gamma_F^{(q)}$;
            \item the set of final states $\overline{Q}_\mathrm{f}$ with $(q,\mathbf{x}_i)\in \overline{Q}_\mathrm{f}\subseteq \mathbb{Q}\times \times (X_\mathrm{var}\cup U_\mathrm{var})$ if there exists $ \mathbf{y}_j\in Y_\mathrm{var}$ such that $(\mathbf{x}_i,\mathbf{y}_j)\in \Gamma_{H}^{(q)}$, and $(q,\mathbf{u}_p)\in \overline{Q}_\mathrm{f}$ if there exists $\mathbf{y}_j\in Y_\mathrm{var}$ such that $(\mathbf{u}_p,\mathbf{y}_j)\in \Gamma_{H}^{(q)}$;
            \item the partial transition function $\delta_\mathcal{H} : \overline{Q} \times (\mathbb{V} \cup \{1\})\to 2^{\overline{Q}}$ is defined as the combination of:
            \begin{enumerate}[label=\roman*)]
                \item the transition relation corresponding to the one-step evolution inside a mode as $(q,\mathbf{x}_j)\in \delta_\mathcal{H}(q,\mathbf{x}_i,1)$ if $(\mathbf{x}_i,\mathbf{x}_j)\in \Gamma_{F}^{(q)}$, or $(q,\mathbf{x}_j)\in \delta_\mathcal{H}(q,\mathbf{u}_p,1)$ if $(\mathbf{u}_p,\mathbf{x}_j)\in \Gamma_{F}^{(q)}$;
                \item the transition relation corresponding to each edge $\eta = (q,q')\in E$ as $(q',\mathbf{x}_i)\in \delta_\mathcal{H}(q,\mathbf{x}_i,w)$ if there exists $w\in \mathbb{V}$.  \hfill $\blacksquare$
				%such that $(\mathbf{x}_i,\mathbf{x}_j)\in \mathcal{R}(\eta,w)$.
            \end{enumerate}
            %The transition relation $\delta_\mathcal{H}(q,\mathbf{x}_i,\sigma) =\emptyset$, $\sigma\in\overline{\Sigma}$, if for $\sigma = \{1\}$ there does not exist $\mathbf{x}_j\in X$ such that $(\mathbf{x}_i,\mathbf{x}_j)\in \Gamma_{F}^{(q)}$ or for $ \sigma = \xi\in \Xi$ there does not exist $q'\in Q$ such that $\xi\in \phi(\eta)$ . 
        \end{itemize}
\end{pf} 
It is noted that the transitions via inputs from $\mathbb{V}$ do not
entail transitions in the state $\mathbf{x}\in X_\mathrm{var}$. Therefore, the
transitions in the mode $q \in \mathbb{Q}$ via $\mathbb{V}$ and one-step state
transitions in $\mathbf{x}\in X_\mathrm{var}$ are allowed to occur
concurrently. Then,
the transition $(q',\mathbf{x}_j)\in \delta_\mathcal{H}(q,\mathbf{x}_i,w)$ for
some $(q,q')\in E$ and $w\in\mathbb{V}$ represents a concatenation of labelled
transitions $(q,\mathbf{x}_i)\xrightarrow{1}(q,\mathbf{x}_j)$ and
$(q,\mathbf{x}_j)\xrightarrow{w}(q',\mathbf{x}_j)$. A similar statement holds
for $(q',\mathbf{x}_j)\in \delta_\mathcal{H}(q,\mathbf{u}_p,w)$.

\section{Model relationships}\label{sec:Relation}
In this section we formalise the relationships between the classes of \Gls{SMPL} models and max-plus automata described in Section \ref{sec:DES-model} and the max-algebraic hybrid automata proposed in Section \ref{sec:MAHA}. To this end, we propose translation procedures among the three modelling classes to further establish partial orders among them.   

%We describe the notions of model containment before proceeding to build relationships between the different modelling classes \cite{Lynch2003}. 

%{\color{red}In order to prove this result, let us first see how any finite-state automaton can always be “transformed” into a Petri net that generates and marks the same languages. Then we will complete the proof by presenting a non-regular language that can be marked by a Petri net.}

\subsection{Pre-order relationships}
We first recall formal notions  from literature for comparison of different modelling classes. This subsection is based entirely on \cite{Polderman1998,Julius2005,VanderSchaft2004}. 

We adopt a behavioural approach towards establishing relationships between different modelling classes, in that the systems are identified as a collection of input-state-output trajectories they allow\footnote{The term \textit{recognised} is usually used instead of \textit{allowed} in automata theory \cite{Cassandras2009}.}.  
\begin{defn}
The behavioural semantics of a dynamical system is defined as a triple $\Omega = (\mathbb{T},\mathbb{S},\mathscr{B})$, where $\mathbb{T}$ is the time axis, $\mathbb{S}$ is the signal space, and $\mathscr{B}\subseteq \mathbb{S}^\mathbb{T}$ is the collection of all possible trajectories allowed by the system. The pair $(\mathbb{T},\mathbb{S})$ is the behavioural \textit{type} of the dynamical system. \hfill $\blacklozenge$
\end{defn}
In the context of this article, $\mathbb{T}=\mathbb{N}$ represents the event counter axis. The signal space $\mathbb{S}$ is factorised as $\mathbb{S} = \mathbb{D}\times\mathbb{I}\times\mathbb{O}$ into the state space $\mathbb{D}$, input space $\mathbb{I}$, and output space $\mathbb{O}$. 
\begin{defn}
	Given a behavioural system model $\Omega = (\mathbb{T},\mathbb{S},\mathscr{B})$ with $\mathbb{S}=\mathbb{D}\times\mathbb{I}\times\mathbb{O}$ factorised into the state, input and output space, respectively. The input-output behaviour of the system model $\Omega$ is the projection of the behaviour $\mathscr{B}$ on the set of input-output signals, $\pi_{\mathrm{IO}}(\mathscr{B}) \subset \mathbb{I}^\mathbb{T}\times \mathbb{O}^\mathbb{T}$. 
\end{defn}

We now proceed to define an input-output behavioural relationship between two dynamical systems.  
\begin{defn}\label{defn:BEq}
	Consider two dynamical systems $\Omega_i=(\mathbb{T},\mathbb{D}_i\times\mathbb{I}\times\mathbb{O},\mathscr{B}_i)$, $i=1,2$. The dynamical system $\Omega_1$ is said to be behaviourally included in $\Omega_2$, denoted as $\Omega_1 \preccurlyeq_\mathrm{B}\Omega_2$, if $\pi_{\mathrm{IO}}(\mathscr{B}_1)\subseteq \pi_{\mathrm{IO}}(\mathscr{B}_2)$. 

	The notion of \textit{behavioural equivalence} (denoted as $\Omega_1\simeq_\mathrm{B}\Omega_2$) follows if the said behavioural inclusion is also symmetric.  \hfill $\blacklozenge$
\end{defn}
The input-output behaviour of a finite automaton can be defined as the collection of all accepted words. In that case, the condition of behavioural equivalence of finite automata implies the equality of their generated languages \cite{Julius2005}.

We now define pre-order relation that also captures the state transitions structures of two dynamical systems. We first define the concept of a state map.
\begin{defn}\label{def:State}
	Given a dynamical system $\Omega=(\mathbb{T},\mathbb{D}\times\mathbb{I}\times\mathbb{O},\mathscr{B})$. A \textit{state map} is defined as a map $\varphi:\mathbb{I}\times\mathbb{O}\times\mathbb{T}\to\mathbb{D}$ such that for every $(x,w,y)\in\mathscr{B}$ there exists $\tau\in\mathbb{T}$ such that $x=\varphi(w,y,\tau)$. \hfill $\blacklozenge$
\end{defn}
The following notion provides a sufficient condition for demonstrating that an input-output behavioural relationship exists between two dynamical systems.  
\begin{defn}\label{def:SimEq}
	Consider two dynamical systems $\Omega_i=(\mathbb{T},\mathbb{D}_i\times\mathbb{I}\times\mathbb{O},\mathscr{B}_i)$, $i=1,2$, and their respective state maps $\varphi_1$ and $\varphi_2$. A \textit{simulation relation} from $\Omega_1$ to $\Omega_2$, $\Psi:\mathbb{T}\to 2^{\mathbb{D}_1\times\mathbb{D}_2}$, is defined such that for any $\tau\in\mathbb{T}$ if $(x_1,x_2)\in\Psi(\tau)$ and $(x_1,w_1,y_1)\in\mathscr{B}_1$ where $x_1 = \varphi_1(w_1,y_1,\tau)$ then there exists $(x_2,w_2,y_2)\in\mathscr{B}_2$ such that $x_2 = \varphi_2(w_2,y_2,\tau)$, and for all $\tau'\geq \tau$ such that $w_1(\tau') = w_2(\tau')$ we have: \rn{1} $(\varphi_1(w_1,y_1,\tau'),\varphi_2(w_1,y_2,\tau'))\in\Psi(\tau')$, and \rn{2} $y_1(\tau') = y_2(\tau')$. 

	The dynamical system $\Omega_1$ is said to be \textit{simulated} by $\Omega_2$, $\Omega_1 \preccurlyeq_\mathrm{S}\Omega_2$, if a simulation relation exists from $\Omega_1$ to $\Omega_2$. 

	The notion of \textit{bisimilarity} (denoted as $\Omega_1\simeq_\mathrm{S}\Omega_2$) follows if the said simulation relation is also symmetric. \hfill $\blacklozenge$
\end{defn}
Finally, we recall the following result from the literature.
\begin{lem}[\cite{Julius2005}]\label{lem:relation}
	Consider two dynamical systems $\Omega_i=(\mathbb{T},\mathbb{D}_i\times\mathbb{I}\times\mathbb{O},\mathscr{B}_i)$, $i=1,2$, and their respective state maps $\varphi_1$ and $\varphi_2$. Then the following implication holds:
	\begin{equation*}
		\Omega_1 \preceq_{\mathrm{S}} \Omega_2\Rightarrow \Omega_1 \preceq_{\mathrm{B}} \Omega_2. \tag*{$\blacksquare$}
	\end{equation*}
\end{lem}
We now move on to formalising the relationships between the proposed max-algebraic hybrid automata and the existing frameworks of \Gls{SMPL} systems and max-plus automata.
\subsection{Equivalent max-algebraic hybrid automata for \Gls{SMPL} systems}
%{\color{blue}The wordings of the following theorems have been changed. The comparisons are now between models and not classes.}

In this subsection we show that \Gls{SMPL} systems in open-loop and closed-loop configurations ($\mathcal{S}_\mathrm{O}$ and $\mathcal{S}_\mathrm{C}$, respectively), are special cases of max-algebraic hybrid automata. To this end, we construct an equivalent restriction of the max-algebraic hybrid automaton. Here, equivalence is expressed in terms of a simulation relation that captures the state transition structure of the \Gls{SMPL} system.
\begin{thm}\label{thm:SO-H}
	Given an open-loop \Gls{SMPL} system $\mathcal{S}_\mathrm{O}$, there exists a max-algebraic hybrid automaton $\mathcal{H}_\mathrm{O}$ that bisimulates it, i.e. $\mathcal{S}_\mathrm{O}\simeq_\mathrm{S}\mathcal{H}_\mathrm{O}$.
\end{thm}
\begin{pf} 
	Consider an open-loop \Gls{SMPL} system $\mathcal{S}_\mathrm{O}$ behaviour consisting of states $(l,x)\in\mathbb{D}=\underline{n_\mathrm{L}}\times\mathbb{R}_\varepsilon^n$, inputs $(w,r)\in\mathbb{B}^{n_\mathrm{L}}_\varepsilon\times\mathbb{R}_\varepsilon^m$, and output $y\in\mathbb{R}_\varepsilon^{d}$ defined on an event counter $k\in\mathbb{N}$. The state maps are defined in \eqref{eq:31} without the control inputs $u$ and $v$ as $x(k) = f(l(k),x(k-1),r(k))$, $l(k)=\phi(l(k),x(k-1),(w(k),r(k)))$ and $y(k) = h(l(k),x(k),r(k))$. The initial condition is denoted as $x_0 = x(0)\in\mathbb{R}^n_\varepsilon$. 
	
	A max-algebraic hybrid automaton $\mathcal{H}_\mathrm{O}$ (as in \eqref{eq:41}) is constructed with the states $q\in \mathbb{Q} = \underline{n_\mathrm{L}}$ and $x_\mathrm{h}\in\mathbb{X}=\mathbb{R}_\varepsilon^n$, the inputs $(w,r)\in\mathbb{I} = \mathbb{V}\times\mathbb{U} = \mathbb{B}^{n_\mathrm{L}}_\varepsilon\times\mathbb{R}_\varepsilon^m$, and the output $y_\mathrm{h}\in\mathbb{Y} = \mathbb{R}_\varepsilon^{d}$. The discrete state characteristics are defined for all $q\in\underline{n_\mathrm{L}}$ as: $(q,x_0)\in\mathrm{Init}$, $F(q,\cdot,\cdot) = f(q,\cdot,\cdot)$, $H(q,\cdot,\cdot) = h(q,\cdot,\cdot)$, and $\mathrm{Inv}(q) = \{(x_\mathrm{h},(w,r))\mid\phi(\cdot,x_\mathrm{h},(w,r))=q\}$. The edge characteristics are defined for all $(q,q')\in E\subseteq \underline{n_\mathrm{L}}\times \underline{n_\mathrm{L}}$ as: $G = \{(x_\mathrm{h},(w,r))\mid\phi(q,x_\mathrm{h},(w,r)) = q'\}$, and $R(\cdot) := R_{\mathrm{id}}(\cdot)$. There are no constraints on the admissible inputs, i.e. $\Lambda(q,x) = 2^{\mathbb{I}}$ for all $(q,x_\mathrm{h})\in\mathbb{X}$.
	% It is assumed that discrete state transitions allowed in $\mathcal{H}_\mathrm{O}$ are instantaneous. 
	%2^{\mathbb{X}\times\mathbb{I}}\cap
	
	Note that the two systems share the same state, input and output spaces. An event counter dependent simulation relation can be defined such that for a given $k'\in\mathbb{N}$, if $((l,x),(q,x_\mathrm{h}))\in\Psi({k'})$ then we have $l(k') = q(k')$ and $x(k') = x_\mathrm{h}(k')$. It is now sufficient to show that the two models produce state trajectories, under the same input sequence $(w(k)),r(k))$ for $k\geq k'$, such that $x(k)=x_\mathrm{h}(k)$ and $l(k)=q(k)$.	
	
	Let $((l,x),(q,x_\mathrm{h}))\in\Psi(k')$, $l(k')=l_1$, and $k''=\inf\{k\in\mathbb{N}\mid \phi(l(k),x(k-1),\cdot)\neq l_1,\; k> k'\}$. We now have that any continuous-valued state trajectory $x(\cdot)$ of the \Gls{SMPL} system inside the mode $l(k)=l_1$, $k\in \{k',k'+1,\dots,k''-1\}$, also satisfies the invariance condition of the mode $q(k)=l_1$. Then, for the same input sequence we have $x_\mathrm{h}(k) = x(k)=f(l_1,\cdot,\cdot)$ as long as $l(k)=q(k)=l_1$. For a mode change $l(k'')=l_2\neq l_1$ such that $\phi(l_1,x(k''-1),\cdot)=l_2$, the invariance condition of mode $q(k)=l_1$ is also violated in the max-algebraic hybrid automaton resulting in a transition in the state from $(l_1,x_\mathrm{h}(k''))$ to $(l_2,x_\mathrm{h}(k''))$ with $x(k'')=x_\mathrm{h}(k'')$. Thus, $((l,x),(q,x_\mathrm{h}))\in\Psi(k)$ for all $k\geq k'$. Moreover, the output function is shared by both the models resulting in $y(k)=y_\mathrm{h}(k)$ for all $k\geq k'$. 
	
	The simulation relation $\Psi(\cdot)$ is indeed symmetric. Hence, we have $\mathcal{S}_\mathrm{O}\simeq_\mathrm{S}\mathcal{H}_\mathrm{O}$.
	%As the model $\mathcal{H}_\mathrm{O}$ is a strict restriction of the model $\mathcal{H}$ (in \eqref{eq:41}), we have $\mathcal{S}_\mathrm{O}\preccurlyeq_\mathrm{S}\mathcal{H}$.
\hfill $\blacksquare$
\end{pf}
\begin{thm}\label{thm:SC-H}
	Given a closed-loop \Gls{SMPL} system $\mathcal{S}_\mathrm{C}$, there exists a max-algebraic hybrid automaton $\mathcal{H}_\mathrm{C}$ that bisimulates it, i.e. $\mathcal{S}_\mathrm{C}\simeq_\mathrm{S}\mathcal{H}$.
\end{thm}
\begin{pf} 
	We now consider a closed-loop \Gls{SMPL} system $\mathcal{S}_\mathrm{C}$ behaviour consisting of states $(l,z)\in\mathbb{D}=\underline{n_\mathrm{L}}\times\mathbb{R}_\varepsilon^{1+n+n_\mathrm{u}+n_\mathrm{v}}$ where $z(k) = [l(k-1),x^\top(k-1),u^\top(k-1),v^\top(k-1)]^\top$, inputs $(w,r)\in\mathbb{B}^{n_\mathrm{L}}_\varepsilon\times\mathbb{R}_\varepsilon^m$, and output $y\in\mathbb{R}_\varepsilon^{d}$ defined on an event counter $k\in\mathbb{N}$. The state maps are defined as compositions of \eqref{eq:31} and \eqref{eq:32} such that $z(k) = f_{\phi,\mathrm{C}}(l(k),z(k-1),r(k))$, $l(k)=\phi(l(k),z(k-1),(w(k),r(k)))$ and $y(k) = h_{\phi,\mathrm{C}}(l(k),z(k),r(k))$.
	%{\color{blue}Note that the control algorithm in \eqref{eq:32} is invariant with respect to the counter $k\in\mathbb{N}$.} 
	The initial condition is denoted as $z_0 = z(0)\in\mathbb{R}^n_\varepsilon$. 
	
	A max-algebraic hybrid automaton $\mathcal{H}_\mathrm{C}$ is constructed with the states $q\in \mathbb{Q} = \underline{n_\mathrm{L}}$ and $x_\mathrm{h}\in\mathbb{X}=\mathbb{R}_\varepsilon^{1+n+n_\mathrm{u}+n_\mathrm{v}}$, the inputs $(w,r)\in\mathbb{I} = \mathbb{V}\times\mathbb{U} = \mathbb{B}^{n_\mathrm{L}}_\varepsilon\times\mathbb{R}_\varepsilon^m$, and the output $y_\mathrm{h}\in\mathbb{Y} = \mathbb{R}_\varepsilon^{d}$. The discrete state characteristics are defined for all $q\in\underline{n_\mathrm{L}}$ as: $(q,z_0)\in\mathrm{Init}$, $F(q,\cdot,\cdot) = f_{\phi,C}(q,\cdot,\cdot)$, $H(q,\cdot,\cdot) = h_{\phi,\mathrm{C}}(q,\cdot,\cdot)$, and $\mathrm{Inv}(q) = \{(x_\mathrm{h},(w,r))\mid\phi(\cdot,x_\mathrm{h},(w,r))=q\}$. The edge characteristics are defined for all $\eta=(q,q')\in E\subseteq \underline{n_\mathrm{L}}\times \underline{n_\mathrm{L}}$ as: $G(\eta) = \{(x_\mathrm{h},(w,r))\mid\phi(q,x_\mathrm{h},(w,r)) = q'\}$, and $R(\cdot) := R_{\mathrm{id}}(\cdot)$. There are no constraints on the admissible inputs, i.e. $\Lambda(q,x) = 2^{\mathbb{I}}$ for all $(q,x_\mathrm{h})\in\mathbb{X}$.
	%2^{\mathbb{X}\times\mathbb{I}}\cap
	
	Then the rest of the proof follows analogously to that of the open-loop case in Theorem \ref{thm:SO-H}. Hence, $\mathcal{S}_\mathrm{C}\simeq_\mathrm{S}\mathcal{H}_\mathrm{C}$.
	\hfill $\blacksquare$
\end{pf}
Due to the findings of the preceding theorem, the discrete transition structure of a max-algebraic hybrid automaton can be classified analogously to the switching mechanism of an \Gls{SMPL} system as presented in Section \ref{sec:switch}.
An open-loop \Gls{SMPL} system with two modes and no continuous-valued inputs is shown in Fig.~\ref{fig:5}.
{\begin{figure}
		\centering
		\includegraphics[width=0.40\textwidth]{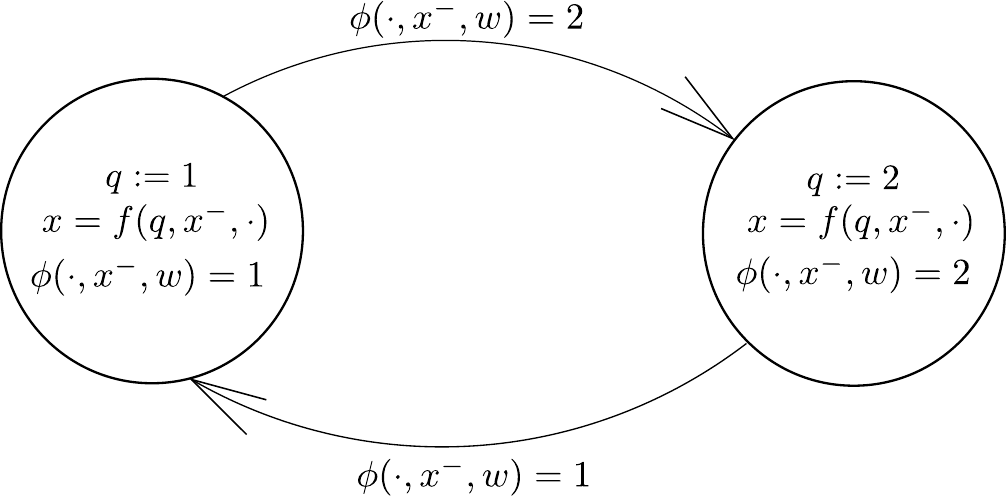}
		\caption{\label{fig:5} A max-algebraic hybrid automaton visualisation of an \Gls{SMPL} system \eqref{eq:31} with $n_\mathrm{L}=2$ modes. The function $\phi(\cdot)$ encoding the switching mechanism appears in the definition of the mode invariants and as directed edge labels specifying the guard set for mode transition. The reset map is identity.}
\end{figure}}   

\subsection{Equivalent max-algebraic hybrid automata for max-plus automata}\label{sec:MAP-SMPL}
This section establishes the relationships between max-plus automata and
max-algebraic hybrid automata. 

We first recall that a max-plus automaton \eqref{eq:39} provides a finite
representation for certain classes of discrete-event systems
\cite{Gaubert1995}. A trajectory of a max-plus automaton $\mathcal{A}$ involves
transitions among discrete states in $S$ such that a (possibly non-unique) accepting path attains the maximum accumulated weight
corresponding to the output \eqref{eq:map-io}. 
%The support of the output sequence,
%$\omega\in\Sigma^*$ such that $y(\omega)$ is finite, is then given as the
%language of the underlying finite automaton $\mathcal{A}_\mathrm{T}$. 
The
auxiliary variable $x(\cdot)$ in \eqref{eq:11}, however, does not constitute the
state space. This is in contrast to the \Gls{SMPL} system description \eqref{eq:31} where the transitions in the hybrid
state $(l,x)$ govern the dynamics. 

We first treat the problem of generating an equivalent max-algebraic hybrid automaton of a given max-plus automaton behaviourally. We show that a subclass of open-loop \Gls{SMPL} systems \eqref{eq:31} generates the same input-output behaviour as that of max-plus automata. The required relationship then follows from the notions presented in the preceding section.
\begin{thm}\label{prop:SO-A}
	Given a max-plus automaton $\mathcal{A}$, there exists an open-loop \Gls{SMPL} system $\mathcal{S}_\mathrm{OA}$ that captures its input-output behaviour, i.e. $\mathcal{A}\preccurlyeq_\mathrm{B}\mathcal{S}_\mathrm{OA}$.
\end{thm}
\begin{pf}
	We first embed the given max-plus automaton
	$\mathcal{A}=(S,\Sigma,\alpha,\mu,\beta)$ into a behavioural model
	consisting of states $s\in\mathbb{D}_1=S=\{s_1,s_2,\dots,s_n\}$, inputs
	$\omega\in\Sigma=\{\sigma_1,\sigma_2,\dots,\sigma_{m}\}$, and output
	$y_a\in\mathbb{R}_\varepsilon$. The input-output behaviour,
	$\pi_\mathrm{IO}(\mathscr{B}_\mathrm{A})$, then consists of the language of
	the max-plus automaton, $\llbracket
	\mathcal{A}\rrbracket_\mathrm{L}\subseteq \Sigma^*$, and the
	output\footnote{Note that with a slight abuse of notation we use the
	shorthand $\boldsymbol{\mu}(\omega_k)=\boldsymbol{\mu}(\gamma_1)\otimes\boldsymbol{\mu}(\gamma_2)\otimes\cdots\otimes\boldsymbol{\mu}(\gamma_k)$.},
	$y_\mathrm{a}(\omega_k)=\boldsymbol{\alpha}^\top\otimes
	\boldsymbol{\mu}(\omega_k)\otimes\boldsymbol{\beta}\in\mathbb{R}$ for
	$\omega_k=\gamma_1\gamma_2\cdots\gamma_k\in\llbracket
	\mathcal{A}\rrbracket_\mathrm{L}$, as in \eqref{eq:11}.  
	
	We recall that the language of the max-plus automaton is a map $\llbracket
	\mathcal{A}\rrbracket_\mathrm{L}: \mathbb{N}\to \Sigma^*$ such that the
	sequence $\omega_k=\gamma_1\gamma_2\cdots\gamma_k\in \Sigma^*$ can be
	represented as a signal $w(j)=\gamma_j$ for all $j\in\underline{k}$. The
	output sequence description can be similarly extended and defined along the
	event counter $k\in\mathbb{N}$. 
	
	Consider an open-loop \Gls{SMPL} system $\mathcal{S}_\mathrm{OA}$ (as in
	\eqref{eq:31}) with the states
	$(l,x)\in\mathbb{D}_2=\underline{m}\times\mathbb{R}_\varepsilon^n$, input
	$w\in \Sigma$, and output $y\in\mathbb{R}_\varepsilon$ defined on an event
	counter $k\in\mathbb{N}$. The state maps (as in \eqref{eq:31}) are defined
	as $x(k) = A^{(l(k))}\otimes x(k-1)$, $l(k)=\phi(\cdot,x(k-1),w(k))$ and
	output as $y(k) = C\otimes x(k)$ where $C\in\mathbb{R}_\varepsilon^{1\times
	n}$, $x(0)\in\mathbb{R}^n_\varepsilon$, $A^{(l)}\in\mathbb{R}^{n\times
	n}_\varepsilon$ for all $l\in\underline{m}$, and 
	\begin{equation}\label{eq:accs}
			\phi(\cdot,x,w) = \left\{l\in\underline{m}\mid A^{(l)}\otimes x\neq \mathcal{E}_{n\times 1},w=\sigma_l\right\}.
	\end{equation}
	For a given initial condition $x(0)\in\mathbb{R}^n_\varepsilon$, the input-output behaviour of the model $\pi_{\mathrm{IO}}(\mathcal{S}_\mathrm{OA})$ consists of input sequences $\{w(k)\}_{k\in\mathbb{N}}$ such that $\phi(\cdot,\cdot,w(k))\neq \emptyset$ and the corresponding output sequences $\{y(k)\}_{k\in\mathbb{N}}$. 
	
	It remains to show that for particular valuations of the matrices $A$ and
	$C$, the max-plus automaton $\mathcal{A}$ and  \Gls{SMPL} system
	$\mathcal{S}_{\mathrm{OA}}$ generate the same input-output behaviour.  

	Consider the specifications: \rn{1} $A^{(l)} = {\mu}^\top(\sigma_l)$ for all
	$l\in\underline{m}$, \rn{2} $[C]_i = \beta(s_i)$, and \rn{3}
	$x_i(0)=\alpha(s_i)$ for $i\in\underline{n}$. Then using \eqref{eq:11},
	given a word $\omega_k=\gamma_1\gamma_2\cdots\gamma_k\in\llbracket\mathcal{A}\rrbracket_\mathrm{L}$ such
	that $w(j)=\gamma_j$, $j\in\underline{k}$, we have $y_\mathrm{a}(\omega_k)=y(k)$ for
	all $k\in\mathbb{N}$. 
	
	Let $x_\mathrm{a}(\cdot)\in\mathbb{R}^{1\times n}_\varepsilon$ denote the auxiliary continuous
	variable satisfying \eqref{eq:11}. Then $x_\mathrm{a}(\omega_j) =
	x_\mathrm{a}(\omega_{j-1})\otimes \boldsymbol{\mu}(\gamma_j)\neq
	\mathcal{E}_{n\times 1}$ for all
	$j\in\underline{k}$ when $\omega_k\in \llbracket\mathcal{A}\rrbracket_\mathrm{L}$. We have
	$x_\mathrm{a}^\top(\omega_{j})= x(j)=A^{(l)}\otimes x(j-1)\neq \mathcal{E}_{n\times
	1}$. Hence, $l\in\phi(\cdot,x(j-1),w(j))$ in \eqref{eq:accs}. Therefore, by
	induction all finite input sequences $\omega_k$ constituting the language of
	the max-plus automaton also satisfy the condition
	$\phi(\cdot,\cdot,w(j))\neq \emptyset$ for $w(j) = \gamma_j$,
	$j\in\underline{k}$. 
	
	Hence, for finite input sequences we have $\pi_\mathrm{IO}(\mathscr{B}_\mathrm{A})\subseteq\pi_\mathrm{IO}(\mathcal{S}_\mathrm{OA})$ resulting in $\mathcal{A}\preccurlyeq_\mathrm{B} \mathcal{S}_\mathrm{OA}$. \hfill $\blacksquare$
\end{pf}
In the preceding proof, we only considered finite input sequences from the input alphabet $\Sigma$. However, the procedure is constructive in that it can be extended to infinite input sequences, by concatenations of finite words from the language $\llbracket\mathcal{A}\rrbracket_\mathrm{L}$, to establish behavioural equivalence. 
{\begin{figure}
		\centering
		\includegraphics[width=0.30\textwidth]{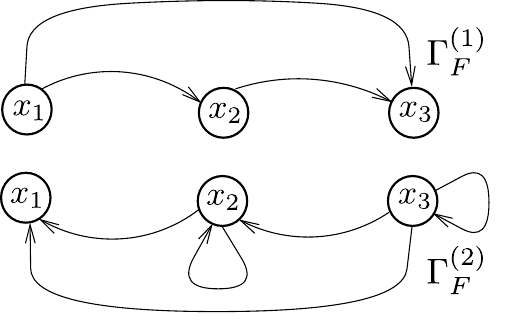}
		\caption{\label{fig:4} The one-step state transition graphs, $\Gamma_{F}^{(1)}$ and $\Gamma_{F}^{(2)}$, as defined in Definition \ref{def:1}, associated to the bimodal open-loop \Gls{SMPL} system of Example \ref{eg:SMPL}.}
\end{figure}}
The above exposition shows that the subclass of discrete-event systems modelled by \Gls{SMPL} systems is at least as large as the subclass modelled by max-plus automata. 

The first relation between max-plus automata and max-algebraic hybrid automata then follows from their respective behavioural relations with SMPL systems. 
%\cbstart
\begin{cor}\label{cor:A-H}
	Given a max-plus automaton $\mathcal{A}$, there exists a max-algebraic hybrid automaton $\mathcal{H}$ (as in \eqref{eq:41}) that captures its input-output behaviour, i.e. $\mathcal{A}\preccurlyeq_\mathrm{B}\mathcal{H}$.
\end{cor}
\begin{pf}
	The proof follows from Lemma \ref{lem:relation}, Theorem \ref{thm:SO-H}, and Theorem \ref{prop:SO-A}.\hfill $\blacksquare$
\end{pf}
\begin{exmp}\label{eg:SMPL}
	Consider an open-loop \Gls{SMPL} system \eqref{eq:31} with three states
	$n=3$, two modes $n_\mathrm{L} = 2$, discrete input
	$w\in\Sigma=\{\sigma_1,\sigma_2\}$ with $\sigma_1=a$ and $\sigma_2=b$. The mode dynamics
	are given for $l\in\underline{n_\mathrm{L}}$:
	\begin{equation}\label{eq:91}
		\begin{aligned}
			f(l,x,\cdot) &= \boldsymbol{\mu}^\top(\sigma_l)\otimes x, &x(0)&=\boldsymbol{\alpha}\\
			h(l,x,\cdot) &= \boldsymbol{\beta}^\top\otimes x,
		\end{aligned}
	\end{equation}
	where $\boldsymbol{\alpha}$, $\boldsymbol{\mu}(\cdot)$ and $\boldsymbol{\beta}$
	are given in \eqref{eq:8}. The underlying one-step state-transition graphs for the mode dynamics, $\Gamma_F^{(l)}$ for $l\in\underline{n_\mathrm{L}}$, are depicted in Fig.~\ref{fig:4}.
	
	The switching function can be obtained from \eqref{eq:accs} for $m = 2$. Then we have, \rn{1} $\left(\boldsymbol{\alpha}^\top\otimes \boldsymbol{\mu}(\sigma_2)\right)^\top=\mathcal{E}_{3\times 1}$, and \rn{2} $\maxpow{\left(\boldsymbol{\mu}(\sigma_1)\right)}{2}\neq\maxpow{\left(\boldsymbol{\mu}(\sigma_1)\right)}{3}=\mathcal{E}_{3\times 1}$. This also means that for discrete inputs with $w(1) = \sigma_2$ and/or $w(k)=w(k+1)=w(k+2)=\sigma_1$ for $k\in\mathbb{N}$, we have $\phi(\cdot,\cdot,w)=\emptyset$.  

	It can now be observed that the described \Gls{SMPL} system is behaviourally equivalent to the max-plus automaton in Example \ref{eg:MPA} following the arguments in Proposition \ref{prop:SO-A}. The max-algebraic hybrid automaton bisimilar to the provided \Gls{SMPL} system is depicted in Fig.~\ref{fig:5}.
\end{exmp}

So far we have established that \Gls{SMPL} systems and, by corollary,
max-algebraic hybrid automata can encode the input-output characteristics of
max-plus automata. We now show that the behaviourally equivalent max-algebraic
hybrid automaton also inherits the state transition (logical) structure of the max-plus automaton. To this end, we consider the finite-state
discrete abstractions of the two systems (as in \eqref{eq:16} and \eqref{eq:13}
respectively) that naturally embed their state transition structure. Then, we
establish a relationship between a max-algebraic hybrid automaton and max-plus
automaton.
\begin{thm}\label{thm:AT-HT}
	Given a max-plus automaton $\mathcal{A}$ with its finite-state discrete
	abstraction denoted as $\mathcal{A}_\mathrm{T}$ (as in \eqref{eq:13}), there
	exists a max-algebraic hybrid automaton $\mathcal{H}$  with a finite-state
	discrete abstraction $\mathcal{H}_\mathrm{OAT}$ (as in Definition
	\ref{prop:HT}) such that $\mathcal{H}_\mathrm{OAT}$ simulates
	$\mathcal{A}_\mathrm{T}$, i.e.
	$\mathcal{A}_\mathrm{T}\preccurlyeq_\mathrm{S}\mathcal{H}_\mathrm{OAT}$.
\end{thm} 
\begin{pf}
	Consider a max-plus automaton $\mathcal{A}=(S,\Sigma,\alpha,\mu,\beta)$ (as
	in \eqref{eq:39}) with state $s\in\mathbb{D}_1=S=\{s_1,s_2,\dots,s_n\}$,
	input $\omega\in\Sigma=\{\sigma_1,\sigma_2,\dots,\sigma_{m}\}$, and output
	$y_a\in\mathbb{R}_\varepsilon$. We recall that the finite-state discrete
	abstraction of the max-plus automaton is a tuple $\mathcal{A}_\mathrm{T} =
	(S,\Sigma,\delta_\mathcal{A},S_\mathrm{0},S_\mathrm{f})$ with \rn{1} a
	partial transition function $\delta_\mathcal{A}:S\times \Sigma\to 2^S$ such
	that $s'\in\delta_\mathcal{A}(s,\sigma)$ if $[\mu(\sigma)]_{ss'}\neq
	\varepsilon$, \rn{2} a set of initial states $S_\mathrm{0}$ such that $s\in
	S_\mathrm{0}$ if $\alpha(s)\neq \varepsilon$, and \rn{3} a set of final
	states $S_\mathrm{f}$ such that $s'\in S_\mathrm{f}$ if $\beta(s')\neq
	\varepsilon$. Moreover, $\llbracket
	\mathcal{A}_\mathrm{T}\rrbracket_\mathrm{L}=\llbracket
	\mathcal{A}\rrbracket_\mathrm{L}$.
	
	We now consider the \Gls{SMPL} system $\mathcal{S}_\mathrm{OA}$ that behaviourally includes the max-plus automaton $\mathcal{A}$ as proposed in Theorem \ref{prop:SO-A}. The max-algebraic hybrid automaton $\mathcal{H}_{\mathrm{OA}}$ such that $\mathcal{S}_\mathrm{OA}\simeq_\mathrm{S} \mathcal{H}_\mathrm{OA}$ can be derived using the procedure described in Theorem \ref{thm:SO-H}. Then $\mathcal{H}_\mathrm{OA}$ consists of \rn{1} states $(q,x)\in \mathbb{Q}\times \mathbb{X} = \underline{m}\times \mathbb{R}^n_\varepsilon$, continuous input $\mathbb{U} = \emptyset$, discrete input $w\in\Sigma$, and $(q,x(0))\in\mathrm{Init}$ for all $q\in \mathbb{Q}$, \rn{2} discrete state characteristics for $x\in\mathbb{R}_\varepsilon^n$ and for all $q\in \mathbb{Q}$ as: $F(q,x,\cdot) = A^{(q)}\otimes x$, $H(q,x,\cdot) = C\otimes x$, and $\mathrm{Inv}(q) = \{(x,w)\mid \phi(\cdot,x,w)\neq \emptyset\}$ (as in \eqref{eq:accs}). The edge characteristics are defined for all $(q,q')\in E\subseteq \underline{n_\mathrm{L}}\times \underline{n_\mathrm{L}}$ as: $G = \{(x,w)\mid\phi(q,x,w) = q'\}$, and $R(\cdot) := R_{\mathrm{id}}(\cdot)$. There are no constraints on the admissible inputs, i.e. $\Lambda(q,x) = 2^{\mathbb{I}}$ for all $(q,x)\in\mathbb{X}$. 

	Now we derive the finite-state discrete abstraction of the max-algebraic hybrid automaton $\mathcal{H}_{\mathrm{OA}}$ following the procedure described in Section \ref{sec:HT}. Recall that the state variables are defined as $X_\mathrm{var}=\{\mathbf{x}_1,\mathbf{x}_2,\dots,\mathbf{x}_n\}$. The transition graphs ($\Gamma_\mathrm{F}^q$ and $\Gamma_\mathrm{H}^q$) for the continuous-variable one-step dynamics (as in Definition \ref{def:1}) reduce to: for all $(i,j)\in \underline{n}^2$ and $q\in \mathbb{Q}$, we have
	\begin{equation}
		\begin{aligned}
			(\mathbf{x}_i,\mathbf{x}_j)\in \Gamma_{\mathrm{F}}^{(q)} &\Leftrightarrow [A^{(q)}]_{ji}\neq \varepsilon,\\
			(\mathbf{x}_j,\mathbf{x}_j)\in \Gamma_{\mathrm{H}}^{(q)} &\Leftrightarrow [C]_j \neq \varepsilon.
		\end{aligned}
	\end{equation}     
	%The reset relation (as in \eqref{eq:19}) can be formulated as $\mathcal{R}:E\times\Sigma\to 2^{{X_\mathrm{var}\times X_\mathrm{var}}}$ such that for every $\eta = (q,q')\in E$ and for every $\sigma\in \Sigma$, we have an identity map for $(i,j)\in \underline{n}^2$ as: $(\mathbf{x}_i,\mathbf{x}_j)\in \mathcal{R}(\eta,\sigma)$ if $\mathbf{x}_i=\mathbf{x}_j$ and there exists $\mathbf{x}_k\in X_\mathrm{var}$ such that $[A^{(q')}]_{kj}\neq \varepsilon$. The latter condition for the reset relation ensures that $\phi(\cdot,\cdot,\sigma)\neq \emptyset$ in \eqref{eq:accs}. 
	%Moreover, Assumption \ref{assum:3} is satisfied. Therefore, the one-step transitions and mode changes via $\Sigma$ can be assumed to occur concurrently (as described in Section \ref{sec:HT}).

	The finite-state discrete abstraction of the max-algebraic hybrid automaton can then be formulated as:
	\begin{equation}\label{eq:HOAT}
		\mathcal{H}_\mathrm{OAT} = (\overline{Q},\Sigma,\delta_\mathcal{H},\overline{Q}_\mathrm{0},\overline{Q}_\mathrm{f}),
	\end{equation}
	where $\overline{Q} = \mathbb{Q}\times X_\mathrm{var}$; $(q,\mathbf{x}_i)\in\overline{Q}_\mathrm{0}$ if $x_i(0)\neq \varepsilon$ and $(q,\mathbf{x}_j)\in\overline{Q}_\mathrm{f}$ if $[C]_j\neq \varepsilon$ for all $q\in Q$; the partial transition function $\delta_\mathcal{H}:\overline{Q}\times \Sigma\to 2^{\overline{Q}}$ is defined such that for $\eta = (q,q')\in E$ and $\sigma\in\Sigma$, we have that $(q',\mathbf{x}_j)\in \delta_\mathcal{H}((q,\mathbf{x}_i),\sigma)$ if $[A^{(q')}]_{ji}\neq \varepsilon$. 

	It remains to show that there exists a simulation relation from $\mathcal{A}_\mathrm{T}$ to $\mathcal{H}_\mathrm{OAT}$ that satisfies the properties stated in Definition \ref{def:SimEq}. The two systems share the same input alphabet $\Sigma$. Moreover, $|\Sigma|=|Q|$ and $|S|=|X_\mathrm{var}|$. Furthermore, $A^{(l)} = \mu^\top(\sigma_l)$ for $l\in\underline{m}$, and $[C]_j = \beta(s_j)$ and $x_j(0) = \alpha(s_j)$ for $j\in\underline{n}$ (as specified in Theorem \ref{prop:SO-A}).

	Recall that words on the input alphabet, $\omega_k=\gamma_1\gamma_2\cdots\gamma_k\in\Sigma^*$, can be identified as a map $\omega:\mathbb{N}\to \Sigma$. Here, $\mathbb{N}$ represents the event counter axis. Also, the partial transition functions, $\delta_\mathcal{A}$ and $\delta_\mathcal{H}$, can be perceived as state maps (as in Definition \ref{def:State}).

	The simulation relation is defined as a map $\Psi:\mathbb{N}\to S\times
	\overline{Q}$ that satisfies the following properties for all
	$k\in\mathbb{N}$: \rn{1} for every $(s_i,(q,\mathbf{x}_j))\in\Psi(k)$ we
	have $i=j$, \rn{2} for every $\sigma_l\in \Sigma$ and
	$(s_i,(q,\mathbf{x}_i))\in \Psi(k)$, we have that for every state
	$s_j\in\{s_t\in\delta_\mathcal{A}(s_i,\sigma_l)\mid
	[\mu(\sigma_l)]_{s_is_t}\neq \varepsilon\}$, there exists
	$(q',\mathbf{x}_j)\in\{(q',x_t)\in
	\delta_\mathcal{H}((q,\mathbf{x}_i),\sigma_l)\mid [A^{(l)}]_{ti}\neq
	\varepsilon\}$ such that $(s_j,(q',\mathbf{x}_j))\in \Psi(k)$, and \rn{3}
	for every $s\in S_\mathrm{0}$ and $(q,\mathbf{x})\in
	\overline{Q}_\mathrm{0}$, we have $(s,(q,\mathbf{x}))\in\Psi(0)$. Note that
	the provided simulation relation is symmetric.  
	
	Therefore, for a given word $\omega_k\in \Sigma^*$ there are equivalent trajectories allowed by $\mathcal{A}_\mathrm{T}$ and $\mathcal{H}_\mathrm{OAT}$. Finally, for every state $s\in\{s_j\in S_\mathrm{f}\mid \beta(s_j)\neq \varepsilon\}$ there exists $(q,\mathbf{x})\in\{(q,\mathbf{x}_j)\in \overline{Q}_\mathrm{f}\mid [C]_j\neq\varepsilon\}$ such that $(s,(q,\mathbf{x}))\in\Psi(k)$, $k\in\mathbb{N}$. Therefore, the final states for the acceptance of the word $\omega_k\in\Sigma^*$ are equivalent in the two models. 

	Hence, we have $\mathcal{A}_{\mathrm{T}}\simeq_\mathrm{S}\mathcal{H}_\mathrm{OAT}$. \hfill $\blacksquare$
\end{pf}
For a max-algebraic hybrid automaton \eqref{eq:41} with max-plus linear mode
dynamics, the finite-state discrete abstraction in \eqref{eq:17} captures
exactly the language of the underlying discrete-event system. The results of the
preceding theorem also imply, using Lemma \ref{lem:relation}, that the two
finite-state discrete abstractions $\mathcal{A}_\mathrm{T}$ and
$\mathcal{H}_\mathrm{OAT}$ and  generate the same language,
$\llbracket\mathcal{A}_\mathrm{T}\rrbracket_\mathrm{L}=\llbracket\mathcal{H}_\mathrm{OAT}\rrbracket_\mathrm{L}$.

% \begin{exmp}
% 	Consider the max-plus automaton in Example \ref{eg:MPA} (depicted in Fig.~\ref{fig:6}) and the behaviourally equivalent \Gls{SMPL} system in Example \ref{eg:SMPL} (and depicted in Fig.~\ref{fig:4}). Then the apparent similarity in the underlying finite automaton \eqref{eq:13} of the max-plus automaton and the underlying directed graphs of the \Gls{SMPL} system is the basis of the proof of Theorem \ref{thm:AT-HT}.
% \end{exmp}
% The practical relevance of the result of this subsection is that the tools designed for analyses and control of max-algebraic hybrid automata can be extended to analyse max-plus automata. 
 
%a discrete-event system specified using both max-algebraic hybrid automata and max-plus automata allow composition based on the proposed translation semantics
\section{Illustration}
\begin{figure}
	\centering
	\includegraphics[width=0.39\textwidth]{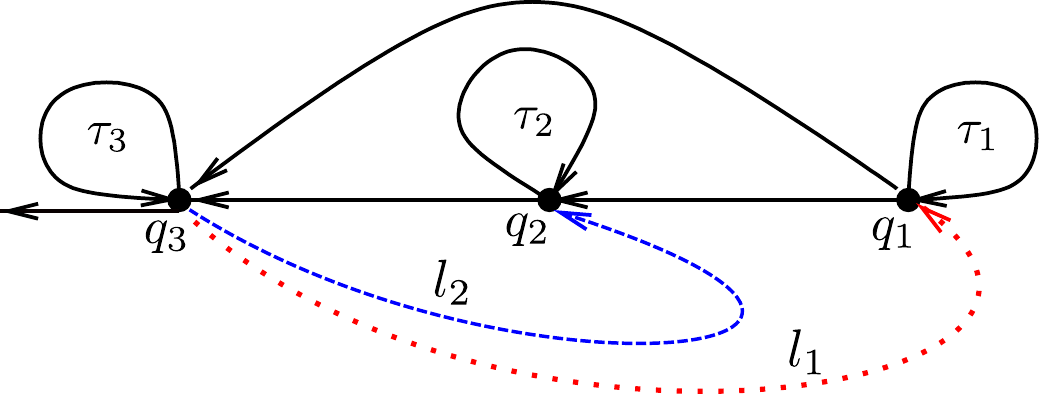}
	\caption{\label{fig:10} A pictorial representation of a production line, adapted from \cite[\S 9.6.1]{Baccelli1992}, \cite[\S 7.2]{SotoyKoelemeijer2003}. The nodes $q_1$-$q_3$ denote machines and are associated with durations $\tau_1$-$\tau_3$ representing processing/recycling times. The two modes of operation can be distinguished by differently coloured arcs: \rn{1} Mode $l_1$ as red dotted line ({\color{red}$\cdots$}), and \rn{2} Mode $l_2$ with blue dashed line ({\color{blue}- - -}).}
\end{figure}

In this subsection, we consider the modelling of a production line, as depicted in Fig. \ref{fig:10}, in the max-algebraic hybrid automata framework. 
%The illustration is adapted from \cite[\S 9.6.1]{Baccelli1992}, \cite[\S 7.2]{SotoyKoelemeijer2003}.

The network consists of nodes $q_1$, $q_2$, and $q_3$ where activities are performed with processing times $\tau_1, \tau_2,\tau_3\in\mathbb{N}$, respectively. The buffers between each pair of nodes have zero holding times and are all assumed to have a single product initially. The buffer before $q_3$ can store at most two incoming products. The other buffers are constrained to hold at most one product at a time. The node $q_1$ transfers product simultaneously to the buffers before $q_2$ and $q_3$. The earliest product\footnote{The conflict at the buffer before $q_3$ is resolved here using the so-called first-in first-out policy.} arriving at $q_3$ is processed first. 

The product exits node $q_3$ and then a new cycle is started. This is modelled as a feedback-loop from node $q_3$ to node $q_1$. In addition, we introduce a second mode of operation where the product from node $q_3$ is routed to node $q_2$ for reprocessing. This is distinguished by differently coloured arcs in Fig~\ref{fig:10}. 

The state $x_i(k)\in\overline{\mathbb{R}}_\varepsilon$, for $i\in\{1,2,3\}$ and $k\in\mathbb{N}$, denotes the time when node $q_i$ finishes an activity for the $k$-th time. The convention is $x_i(k)=+\infty$ if no activity is performed at $q_i$ for the $k$-th time. It is assumed that all buffers contain a product initially. The dynamics of the production line can be expressed algebraically (as in \eqref{eq:31}) as follows for mode $\ell(\cdot) = l_1$:
\begin{equation} \label{eq:6}
    \begin{aligned}
%		x(k+1) &= \min{j\in \underline{L}}\; A^{(i,j)}\otimes x(k),
%		y(k+1) &= C^{(i)}\otimes x(k).
        x_1(k+1) &=\max \left(x_1(k)+\tau_{1}, x_2(k), x_3(k)+\tau_3\right) \\
        x_2(k+1) &=\max \left(x_1(k)+\tau_{1}, x_2(k)+\tau_{2}\right) \\
        x_3(k+1) &=\begin{aligned}[t]\max (&x_1(k)+\tau_1, x_2(k)+\tau_{2},x_3(k)+2\tau_3,\\ &\min(x_1(k)+\tau_1+\tau_3,x_2(k)+\tau_2+\tau_3)).\end{aligned}
        \end{aligned}
\end{equation}
%where
%\begin{equation}
%	\begin{aligned}
%		A^{(l_1,1)} = \begin{pmatrix}
%			\tau_a& 0 & \varepsilon & 0\\
%			\tau_a& \tau_b & \varepsilon & \varepsilon\\
%			\tau_a& \varepsilon & \varepsilon & \tau_d\\
%			\tau_a& \tau_b& 
%		\end{pmatrix}
%	\end{aligned}
%\end{equation}
For the system dynamics in mode $\ell(\cdot)= l_2$, we have:
\begin{equation}\label{eq:7}
    \begin{aligned}
        x_1(k+1) &= \max \left(x_1(k)+\tau_{1}, x_2(k)\right)\\
        x_2(k+1) &=\max \left(x_1(k)+\tau_{1}, x_2(k)+\tau_{2},x_3(k)+\tau_3\right), 
        \end{aligned}
\end{equation}
and the evolution of $x_3$ follows the same equation as of mode $l_1$. The initial state and output matrices ($y=C\otimes x$) are chosen as follows:
\begin{equation}
	\begin{aligned}
		x(0) &= \begin{pmatrix}0 &0 &\varepsilon\end{pmatrix}^\top,\quad	
		C = \begin{pmatrix}\varepsilon &\varepsilon &0\end{pmatrix}.
	\end{aligned}
\end{equation}
The dynamics can be represented in the min-max-plus conjunctive normal form \eqref{eq:4}, for $L=2$ and $M=1$, by replacing the expression of $x_3(\cdot)$ in \eqref{eq:6} with
\begin{equation}
	x_3(k+1)  =\begin{aligned}[t]
		\min \{&\begin{aligned}[t]\max (x_1(k)+\tau_1+\tau_3, x_2&(k)+\tau_2,\\ &x_3(k)+2\tau_3),\end{aligned}\\ & \begin{aligned}[t]\max (x_1(k)+\tau_1, x_2(k)+&\tau_2+\tau_3,\\ &x_3(k)+2\tau_3)\}.\end{aligned}
	\end{aligned}
\end{equation}
There are no continuous-valued inputs to the system. The discrete input $w(\cdot)\in\mathbb{V}\triangleq \{l_1,l_2\}$ determines the mode as follows (see \eqref{eq:31}):
\begin{equation}\label{eq:accs1}
	\begin{aligned}
		\phi(\cdot,x,w) = \bigg\{i\in\{1,2\}\;\biggl\vert\; \min_{j\in \underline{L}} A^{(i,j)}\otimes x\in \overline{\mathbb{R}}^n_\varepsilon\setminus &\{\varepsilon,\top\}^n ,\\ &w=l_i\bigg\}.
	\end{aligned}
\end{equation}
The discrete-event system of the production network under consideration can therefore be expressed as a max-algebraic hybrid automaton as depicted in Fig.~\ref{fig:5} with continuous-valued dynamics of the form \eqref{eq:4}. 

\begin{figure}
	\centering
	\includegraphics[width=0.39\textwidth]{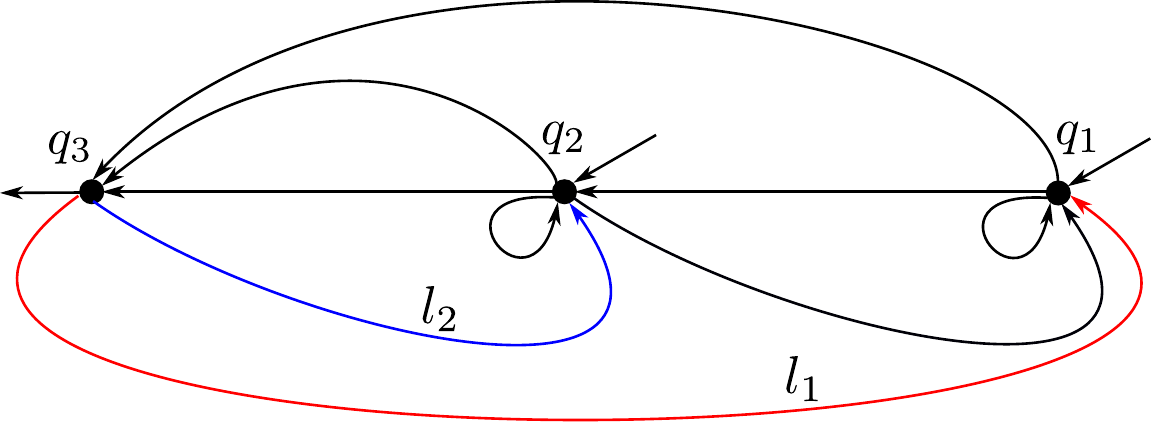}
	\caption{\label{fig:20} The one-step state transition graphs associated to the production network in Fig. \ref{fig:10}. The finite automaton can be obtained by duplication of the nodes $q_1-q_3$ for the two modes $l_1$ and $l_2$. The black arcs are common to both the modes. The blue arc ({\color{blue}---}) belongs to mode $l_2$ and the red arcs ({\color{red}---}) belong to mode $l_1$. The input and output arrows symbolise the initial and final states of the finite automaton.}
\end{figure}

As the system dynamics \eqref{eq:6}-\eqref{eq:7} satisfy Assumption
\ref{assum:2}, a finite-state discrete abstraction of the max-algebraic hybrid
automaton can be obtained using Proposition~\ref{prop:HT}. The necessity of the
restriction of the state space $\mathbb{X}$ is reflected in the definition of
the switching function $\phi(\cdot)$ in \eqref{eq:accs1}. The resulting one-step
state transition graphs of the two modes are depicted in Fig.~\ref{fig:20}.
Moreover, the reset relation does not entail transitions in continuous-valued
state. Then the language of the max-algebraic hybrid automaton model of the
production network is contained in the language of the obtained finite
automaton.
%\begin{multicols}{2}
This completes the illustration.

\section{Conclusions}
In this article, we have proposed a unifying max-algebraic hybrid automata
framework for discrete-event systems in max-plus algebra. In this context, we
identify the hybrid phenomena due to the interaction of continuous-valued
max-plus dynamics and discrete-valued switching dynamics in switching max-plus
linear and max-plus automata models. We have formally established the
relationship between these two models and their relationships with the proposed
max-algebraic hybrid automata framework utilising the notions of behavioural
equivalence and bisimilarity. This is achieved in a behavioural framework where
the models are seen as a collection of input-state-output trajectories. As a
max-algebraic hybrid automaton and a max-plus automaton are defined on different
state space, we have also studied their relationship by embedding them into
their respective finite-state discrete abstractions. 

In the future, we would like to identify the subclass of max-algebraic hybrid automata that can be simulated by a max-plus automaton. We would also like to address the relationships among timed Petri nets, extensions of max-plus automata and max-algebraic hybrid automata. 
%further develop and access the potential of the novel framework presented in this article. In particular, we will
%Finally, we would like to extend the notions and tools of stability and controllability of max-plus linear systems to max-algebraic hybrid automata.  
\bibliography{library1}

\begin{thebibliography}{10}

\bibitem{Baccelli1992}
F.~Baccelli, G.~Cohen, G.~J. Olsder, and J.-P. Quadrat.
\newblock {\em Synchronization and Linearity: An Algebra for Discrete Event
  Systems}.
\newblock John Wiley {\&} Sons, 1992.

\bibitem{Cassandras2009}
C.~G. Cassandras and S.~Lafortune.
\newblock {\em Introduction to Discrete Event Systems}.
\newblock Springer Science {\&} Business Media, 2009.

\bibitem{Cohen1997a}
G.~Cohen, S.~Gaubert, and J.-P. Quadrat.
\newblock Algebraic system analysis of timed {P}etri nets.
\newblock In {\em Idempotency}, pages 145--170. Cambridge University Press,
  1997.

\bibitem{Cohen1999}
G.~Cohen, S.~Gaubert, and J.-P. Quadrat.
\newblock Max-plus algebra and system theory: Where we are and where to go now.
\newblock {\em Annual Reviews in Control}, 23:207--219, Jan. 1999.

\bibitem{DeSchutter2001}
B.~{De Schutter} and T.~J. van~den Boom.
\newblock Model predictive control for max-plus-linear discrete event systems.
\newblock {\em Automatica}, 37(7):1049--1056, July 2001.

\bibitem{DeSchutter2004a}
B.~{De Schutter} and T.~J. van~den Boom.
\newblock {MPC} for continuous piecewise-affine systems.
\newblock {\em Systems and Control Letters}, 52(3-4):179--192, July 2004.

\bibitem{Gaubert1995}
S.~Gaubert.
\newblock Performance evaluation of (max,+) automata.
\newblock {\em IEEE Transactions on Automatic Control}, 40(12):2014--2025,
  1995.

\bibitem{Gaubert1999}
S.~Gaubert and J.~Mairesse.
\newblock Modeling and analysis of timed {P}etri nets using heaps of pieces.
\newblock {\em IEEE Transactions on Automatic Control}, 44(4):683--697, 1999.

\bibitem{Gunawardena1994}
J.~Gunawardena.
\newblock Min-max functions.
\newblock {\em Discrete Event Dynamic Systems: Theory and Applications},
  4(4):377--407, 1994.

\bibitem{Heemels2001}
W.~P. M.~H. Heemels, B.~{De Schutter}, and A.~Bemporad.
\newblock Equivalence of hybrid dynamical models.
\newblock {\em Automatica}, 37(7):1085--1091, July 2001.

\bibitem{heidergott2014max}
B.~Heidergott, G.~J. Olsder, and J.~van~der Woude.
\newblock {\em Max Plus at Work: Modeling and Analysis of Synchronized Systems:
  A Course on Max-Plus Algebra and its Applications}.
\newblock Princeton University Press, 2014.

\bibitem{Julius2005}
A.~A. Julius and A.~J. van~der Schaft.
\newblock Bisimulation as congruence in the behavioral setting.
\newblock In {\em Proceedings of the 44th IEEE Conference on Decision and
  Control}, pages 814--819, 2005.

\bibitem{Komenda2018}
J.~Komenda, S.~Lahaye, J.~L. Boimond, and T.~J. van~den Boom.
\newblock Max-plus algebra in the history of discrete event systems.
\newblock {\em Annual Reviews in Control}, 45:240--249, Jan. 2018.

\bibitem{Lahaye2008}
S.~Lahaye, J.-L. Boimond, and J.-L. Ferrier.
\newblock Just-in-time control of time-varying discrete event dynamic systems
  in (max,+) algebra.
\newblock {\em International Journal of Production Research},
  46(19):5337--5348, 2008.

\bibitem{Lygeros1998}
J.~Lygeros, D.~N. Godbole, and S.~Sastry.
\newblock Verified hybrid controllers for automated vehicles.
\newblock {\em IEEE Transactions on Automatic Control}, 43(4):522--539, 1998.

\bibitem{Lygeros1999}
J.~Lygeros, G.~Pappas, and S.~Sastry.
\newblock An introduction to hybrid system modeling, analysis, and control.
\newblock In {\em Preprints of the First Nonlinear Control Network Pedagogical
  School}, pages 1--14, 1999.

\bibitem{Lygeros1999a}
J.~Lygeros, C.~Tomlin, and S.~Sastry.
\newblock Controllers for reachability specifications for hybrid systems.
\newblock {\em Automatica}, 35(3):349--370, Mar. 1999.

\bibitem{Maia2003}
C.~A. Maia, L.~Hardouin, R.~Santos-Mendes, and B.~Cottenceau.
\newblock Optimal closed-loop control of timed event graphs in dioids.
\newblock {\em IEEE Transactions on Automatic Control}, 48(12):2284--2287, Dec.
  2003.

\bibitem{Olsder1991}
G.~J. Olsder.
\newblock Eigenvalues of dynamic max-min systems.
\newblock {\em Discrete Event Dynamic Systems: Theory and Applications},
  1(2):177--207, Sept. 1991.

\bibitem{SotoyKoelemeijer2003}
G.~{Soto Y Koelemeijer}.
\newblock {\em On the Behaviour of Classes of Min-Max-Plus Systems}.
\newblock PhD thesis, Delft University of Technology, 2003.

\bibitem{Torrisi2004}
F.~D. Torrisi and A.~Bemporad.
\newblock {HYSDEL} - {A} tool for generating computational hybrid models for
  analysis and synthesis problems.
\newblock {\em IEEE Transactions on Control Systems Technology},
  12(2):235--249, Mar. 2004.

\bibitem{VanDenBoom2002a}
T.~J. van~den Boom and B.~{De Schutter}.
\newblock Properties of {MPC} for max-plus-linear systems.
\newblock {\em European Journal of Control}, 8(5):453--462, Jan. 2002.

\bibitem{VandenBoom2006}
T.~J. van~den Boom and B.~{De Schutter}.
\newblock Modelling and control of discrete event systems using switching
  max-plus-linear systems.
\newblock {\em Control Engineering Practice}, 14(10):1199--1211, Oct. 2006.

\bibitem{VanDenBoom2012b}
T.~J. van~den Boom and B.~{De Schutter}.
\newblock Modeling and control of switching max-plus-linear systems with random
  and deterministic switching.
\newblock {\em Discrete Event Dynamic Systems: Theory and Applications},
  22(3):293--332, Sept. 2012.

\bibitem{VandenBoom2020}
T.~J. van~den Boom, M.~van~den Muijsenberg, and B.~{De Schutter}.
\newblock Model predictive scheduling of semi-cyclic discrete-event systems
  using switching max-plus linear models and dynamic graphs.
\newblock {\em Discrete Event Dynamic Systems}, 30(4):1--35, 2020.

\bibitem{VanderSchaft2004}
A.~J. van~der Schaft.
\newblock Equivalence of dynamical systems by bisimulation.
\newblock {\em IEEE Transactions on Automatic Control}, 49d(12):2160--2172,
  Dec. 2004.

\bibitem{Polderman1998}
J.~C. Willems and J.~W. Polderman.
\newblock {\em Introduction to Mathematical Systems Theory: A Behavioral
  Approach}.
\newblock Springer-Verlag, New York, 1998.

\end{thebibliography}

\end{document}